\documentclass[11pt]{article}

\usepackage{amsmath,amssymb}
\usepackage{rotating}
\usepackage{array}
\usepackage{epsfig}

\usepackage{amsmath}
\usepackage{graphicx}
\usepackage{latexsym}

\newcount\figureno     \figureno=0
\newdimen\figdim       \figdim=70mm
\def\figureinc{%
   \global\advance\figureno by 1%
}
\def\figcaption#1#2#3{\hbox to #2{\hss{\vbox{\hsize=#2 \parindent=0pt
        {\bf Figure \number\figureno#3 :\ }#1}}\hss}
}

\begin{document}
\baselineskip 100pt

{\large
\parskip.2in
\newcommand{\be}{\begin{equation}}
\newcommand{\ee}{\end{equation}}
\newcommand{\ben}{\begin{equation*}}
\newcommand{\een}{\end{equation*}}
\newcommand{\br}{\bar}
\newcommand{\fr}{\frac}
\newcommand{\lm}{\lambda}
\newcommand{\ra}{\rightarrow}
\newcommand{\al}{\alpha}
\newcommand{\bt}{\beta}
\newcommand{\z}{\zeta}
\newcommand{\pa}{\partial}
\newcommand{\hs}{\hspace{5mm}}
\newcommand{\up}{\upsilon}
\newcommand{\dg}{\dagger}
\newcommand{\sdil}{\ensuremath{\rlap{\raisebox{.15ex}{$\mskip
6.5mu\scriptstyle+ $}}\subset}}
\newcommand{\sdir}{\ensuremath{\rlap{\raisebox{.15ex}{$\mskip
6.5mu\scriptstyle+ $}}\supset}}
\newcommand{\vphi}{\vec{\varphi}}
\newcommand{\ve}{\varepsilon}
\newcommand{\acc}{\\[3mm]}
\newcommand{\dl}{\delta}
\def\tablecap#1{\vskip 3mm \centerline{#1}\vskip 5mm}
\def\p#1{\partial_#1}
\newcommand{\pd}[2]{\frac{\partial #1}{\partial #2}}
\newcommand{\pdn}[3]{\frac{\partial #1^{#3}}{\partial #2^{#3}}}
\def\DP#1#2{D_{#1}\varphi^{#2}}
\def\dP#1#2{\partial_{#1}\varphi^{#2}}
\def\xh{\hat x}
\newcommand{\Ref}[1]{(\ref{#1})}

\def\mod#1{ \vert #1 \vert }
\def\chapter#1{\hbox{Introduction.}}
\def\Sin{\hbox{sin}}
\def\Cos{\hbox{cos}}
\def\Exp{\hbox{exp}}
\def\Ln{\hbox{ln}}
\def\Tan{\hbox{tan}}
\def\Cot{\hbox{cot}}
\def\Sinh{\hbox{sinh}}
\def\Cosh{\hbox{cosh}}
\def\Tanh{\hbox{tanh}}
\def\Asin{\hbox{asin}}
\def\Acos{\hbox{acos}}
\def\Atan{\hbox{atan}}
\def\Asinh{\hbox{asinh}}
\def\Acosh{\hbox{acosh}}
\def\Atanh{\hbox{atanh}}
\def\frac#1#2{{\textstyle{#1\over #2}}}

\def\ph{\varphi_{m,n}}
\def\phl{\varphi_{m-1,n}}
\def\phr{\varphi_{m+1,n}}
\def\varphil{\varphi_{m-1,n}}
\def\varphir{\varphi_{m+1,n}}
\def\varphit{\varphi_{m,n+1}}
\def\varphib{\varphi_{m,n-1}}
\def\pht{\varphi_{m,n+1}}
\def\phb{\varphi_{m,n-1}}
\def\phbl{\varphi_{m-1,n-1}}
\def\phbr{\varphi_{m+1,n-1}}
\def\phtl{\varphi_{m-1,n+1}}
\def\phtr{\varphi_{m+1,n+1}}
\def\u{u_{m,n}}
\def\ul{u_{m-1,n}}
\def\ur{u_{m+1,n}}
\def\ut{u_{m,n+1}}
\def\ub{u_{m,n-1}}
\def\utr{u_{m+1,n+1}}
\def\ubl{u_{m-1,n-1}}
\def\utl{u_{m-1,n+1}}
\def\ubr{u_{m+1,n-1}}
\def\v{v_{m,n}}
\def\vl{v_{m-1,n}}
\def\vr{v_{m+1,n}}
\def\vt{v_{m,n+1}}
\def\vb{v_{m,n-1}}
\def\vtr{v_{m+1,n+1}}
\def\vbl{v_{m-1,n-1}}
\def\vtl{v_{m-1,n+1}}
\def\vbr{v_{m+1,n-1}}

\def\U{U_{m,n}}
\def\Ul{U_{m-1,n}}
\def\Ur{U_{m+1,n}}
\def\Ut{U_{m,n+1}}
\def\Ub{U_{m,n-1}}
\def\Utr{U_{m+1,n+1}}
\def\Ubl{U_{m-1,n-1}}
\def\Utl{U_{m-1,n+1}}
\def\Ubr{U_{m+1,n-1}}
\def\V{V_{m,n}}
\def\Vl{V_{m-1,n}}
\def\Vr{V_{m+1,n}}
\def\Vt{V_{m,n+1}}
\def\Vb{V_{m,n-1}}
\def\Vtr{V_{m+1,n+1}}
\def\Vbl{V_{m-1,n-1}}
\def\Vtl{V_{m-1,n+1}}
\def\Vbr{V_{m+1,n-1}}

\newcommand{\ie}{{\it i.e.}}
\newcommand{\cmod}[1]{ \vert #1 \vert ^2 }
\newcommand{\cmodn}[2]{ \vert #1 \vert ^{#2} }
\newcommand{\nhat}{\mbox{\boldmath$\hat n$}}
\nopagebreak[3]
\bigskip

\title{ \bf $N=2$ supersymmetric extension of a hydrodynamic system in Riemann invariants }
\vskip 1cm

\bigskip
\author{
A.~M. Grundland\thanks{email address: grundlan@crm.umontreal.ca}
\\
Centre de Recherches Math{\'e}matiques, Universit{\'e} de Montr{\'e}al,\\
C. P. 6128, Succ.\ Centre-ville, Montr{\'e}al, (QC) H3C 3J7,
Canada\\ Universit\'{e} du Qu\'{e}bec, Trois-Rivi\`{e}res, CP500 (QC) G9A 5H7, Canada \acc A. J. Hariton\thanks{email address: hariton@crm.umontreal.ca}
\\
Centre de Recherches Math{\'e}matiques, Universit{\'e} de Montr{\'e}al, \\
C. P. 6128, Succ.\ Centre-ville, Montr{\'e}al, (QC) H3C 3J7, Canada \\} \date{}

\maketitle

\begin{abstract}

In this paper, we formulate an $N=2$ supersymmetric extension of a hydrodynamic-type system involving Riemann invariants. The supersymmetric version is constructed by means of a superspace and superfield formalism, using bosonic superfields, and consists of a system of partial differential equations involving both bosonic and fermionic variables. We make use of group-theoretical methods in order to analyze the extended model algebraically. Specifically, we calculate a Lie superalgebra of symmetries of our supersymmetric model and make use of a general classification method to classify the one-dimensional subalgebras into conjugacy classes. As a result we obtain a set of 401 one-dimensional nonequivalent subalgebras. For selected subalgebras, we use the symmetry reduction method applied to Grassmann-valued equations in order to determine analytic exact solutions of our supersymmetric model. These solutions include travelling waves, bumps, kinks, double-periodic solutions and solutions involving exponentials and radicals.

\end{abstract}




\newpage

\section{Introduction}

The concept of supersymmetry has been of considerable interest in the past three decades. Supersymmetric theories involving both bosonic and fermionic degrees of freedom have been used extensively in order to describe various types of physical phenomena. We now list some of the most important examples. Progress in the analytic treatment of supersymmetric extensions of existing  models, first used in the context of particle physics, and later in classical field theories such as fluid mechanics, has been rapid and resulted in many new techniques and theoretical approaches \cite{Manin,Fatyga,Roelofs}. Some of the most interesting developments have been in the study of supersymmetric extensions for fluid dynamics, including polytropic gas dynamics \cite{Das}, relativistic hydrodynamics in four-dimensional Minkowski space \cite{Nyawelo1,Nyawelo2}, and a Kaluza-Klein model of a relativistic fluid \cite{Hassaine}. More recently, the techniques have been applied to sigma models. The quantum version of the $\mathbb{C}P^{n-1}$ model contains an anomaly which prevents its integrability. However, in the $N=1$ supersymmetric version of the $\mathbb{C}P^{n-1}$ model (involving only one fermionic independent variable), this anomaly disappears and the quantum version becomes an integrable model \cite{Abdalla}.

In the literature, there exist several ways of supersymmetrizing nonlinear models. For instance, supersymmetric extensions of the Chaplygin gas in $(1+1)$ and $(2+1)$ dimensions were formulated through parametrizations of the Nambu-goto superstring and supermembrane \cite{Jackiw,Bergner,Polychronakos}. The approach used in this paper involves a formalism in which the space of independent variables is enlarged to a superspace involving both bosonic and fermionic variables, while the dependent fields are replaced by generalized superfields. In particular, this method was used in the case of the Korteweg-de Vries equation \cite{Mathieu,Labelle} and a version of the symmetry reduction method has been extended to the case involving Grassmann-valued partial differential equations in order to determine its invariant solutions \cite{HusAyaWin}.

It is well known \cite{Raif1,Niederer,LevyLeblond1,Belavin} that there exists a connection between the Lie point symmetries of the one-component Schr\"{o}dinger equation for a spinless particle
and those of hydrodynamic systems at the classical level. The Lie algebra of infinitesimal symmetries of the one-component Schr\"{o}dinger equation in $d$ dimensions (denoted $Sch_{4+d(d+3)/2}$) is spanned by the following vector fields
\begin{equation}
\begin{split}
& H=\partial_t,\qquad P^i=\partial_{x^i},\qquad L^{ij}=\frac{1}{2}\left(x^i\partial_{x^j}-x^j\partial_{x^i}\right),\qquad M=\partial_u\\ & G^i=t\partial_{x^i}+x^i\partial_u,\qquad D=t\partial_t+\frac{1}{2}\sum_{i=1}^dx^i\partial_{x^i},\qquad F=\frac{1}{2}t^2\partial_t+\frac{1}{2}\sum_{i=1}^dtx^i\partial_{x^i}+\frac{1}{4}x^2\partial_u.
\end{split}
\label{SchrSymm}
\end{equation}
It was demonstrated \cite{Takahashi1,Takahashi2} that a dynamical system whose conservation laws are the same as for a Schr\"{o}dinger free particle must have a Lagrangian density of the form
\begin{equation}
{\mathcal L}=\left[-\left(u_t+\frac{1}{2}\sum_{i=1}^d(\partial_iu)^2\right)\right]^{1+d/2}.
\label{lagrangianschr}
\end{equation}
If we introduce the vector $\mathbf{v}$ such that $v^i=\partial_iu$ and the density function $\rho=-\partial{\mathcal L}/\partial u_t$, then the potential $u$ can be eliminated from the equations of motion, which become
\begin{equation}
\mathbf{v}_t+(\mathbf{v}\cdot\mathbf{\nabla})\mathbf{v}+{1\over \rho}\mathbf{\nabla}P=0,\qquad \rho_t+\mathbf{\nabla}\cdot(\rho\mathbf{v})=0,
\label{arrow1}
\end{equation}
and are identified as the equations of fluid dynamics describing the potential isentropic flow of a gas, where the (polytropic) pressure is $P(\rho)=A\rho^{1+d/2}$, $A\in\mathbb{R}$. This system can be easily solved using the method of characteristics leading to the non-scattering double waves
\begin{equation}
\begin{split}
u=k^{1/2}(r^1-r^2)+u_0,\quad\rho=Ae^{r^1+r^2},\quad p=kAe^{r^1+r^2}+p_0,\quad u_0,p_0,k,A\in\mathbb{R},
\end{split}
\label{a2}
\end{equation}
which reduces (\ref{arrow1}) to the following invariant hydrodynamic system
\begin{equation}
\begin{split}
r^1_t+\left(k^{1/2}(r^1-r^2+1)+u_0\right)r^1_x=0,\qquad
r^2_t+\left(k^{1/2}(r^1-r^2-1)+u_0\right)r^2_x=0.
\end{split}
\label{a3}
\end{equation}
The asymptotic behavior of initial localized disturbances (i.e. initial data with compact support) corresponding to the waves described in (\ref{a2})  was studied by Riemann \cite{Riemann2}. It was demonstrated that after some finite time $T$, these waves could be separated again in such a way that waves of the same type as those assumed in the initial data could be observed. After a certain finite time $T$ the first derivatives of the solution become unbounded so that, for times $t>T$, smooth solutions to the Cauchy problem do not exist. This phenomenon is known as the gradient catastrophe \cite{Courant1,John}. Invariant solutions of the Schr\"{o}dinger equation display similar behavior \cite{Gagnon}.

As a subcase of the hydrodynamic model (\ref{arrow1}), we consider the equations of a steady, irrotational and compressible fluid flow in a plane \cite{Loewner}
\begin{equation}
\begin{split}
u_y-v_x&=0,\qquad
(\rho u)_x+(\rho v)_y=0,
\end{split}
\label{c1}
\end{equation}
where $(u,v)$ are the Cartesian components of the fluid velocity expressed in terms of a velocity potential $\varphi$ (where $u=\varphi_x$, $v=\varphi_y$), and the density $\rho$ is assumed to be a function of $u$ and $v$ only. Different choices of function for the density lead to different hydrodynamic models. Two of the most important such choices are a Gaussian, irrotational, compressible fluid flow
\begin{equation}
\left(1-(\varphi_x)^2\right)\varphi_{xx}-2\varphi_x\varphi_y\varphi_{xy}+\left(1-(\varphi_y)^2\right)\varphi_{yy}=0,
\label{poom1}
\end{equation}
in which case the density is given by the state equation
\begin{equation}
\rho=e^{-u^2-v^2},
\label{c2}
\end{equation}
and the minimal surfaces equation in $(2+1)$-dimensional Minkowski space
\begin{equation}
\left(\varphi_x\over (1+(\varphi_x)^2+(\varphi_y)^2)^{1/2}\right)_x+\left(\varphi_y\over (1+(\varphi_x)^2+(\varphi_y)^2)^{1/2}\right)_y=0,
\label{poom2}
\end{equation}
corresponding to the state equation
\begin{equation}
\rho=(1+u^2+v^2)^{-1/2}.
\label{c3}
\end{equation}
Through the use of the Wick rotation $y=it$, equation (\ref{poom2}) can be connected to the scalar Born-Infeld equation
\begin{equation}
\left(1+(\varphi_x)^2\right)\varphi_{tt}-2\varphi_x\varphi_t\varphi_{xt}-\left(1-(\varphi_t)^2\right)\varphi_{xx}=0.
\label{c5}
\end{equation}
The Born-Infeld equation (\ref{c5}) is compatible with the following hydrodynamic type system expressed in terms of Riemann invariants \cite{Jackiw}
\begin{equation}
R_t+SR_x=0,\qquad S_t+RS_x=0,
\label{b16}
\end{equation}
via the transformation 
\begin{equation}
\begin{split}
& R=-{(1+(\varphi_x)^2-(\varphi_t)^2)^{1/2}\over 1+(\varphi_x)^2}-{\varphi_x\varphi_t\over 1+(\varphi_x)^2},\\
& S={(1+(\varphi_x)^2-(\varphi_t)^2)^{1/2}\over 1+(\varphi_x)^2}-{\varphi_x\varphi_t\over 1+(\varphi_x)^2}.
\end{split}
\label{c7}
\end{equation}
Finally, the system for the Chaplygin gas
\begin{equation}
U_t=\frac{1}{2}\left(U^2-V^{-2}\right)_x,\qquad
V_t=\left(UV\right)_x,
\label{qwerty10}
\end{equation}
expressed as a conservation law, can be linked to the system (\ref{b16}) in Riemann invariants through the relations
\begin{equation}
R=U+{1\over V},\qquad S=U-{1\over V}.
\label{qwerty11}
\end{equation}
For both the scalar Born-Infeld equation (\ref{c5}) and the Gaussian fluid flow (\ref{poom1}), $N=1$ supersymmetric generalizations were formulated through the use of a superspace and superfield formalism \cite{Hariton4, GrundHaritGauss}. A group-theoretical analysis of a supersymmetric hydrodynamic-type system allows us to characterize a Lie superalgebra of infinitesimal symmetries of the extended system and perform a systematic classification of its subalgebras. The method of symmetry reduction for Grassmann-valued equations allows us to determine exact analytic solutions of our generalized supersymmetric model.
In a previous article \cite{GrundHaritRiemann}, we constructed an $N=1$ extension of the hydrodynamic type system (\ref{b16})
through the use of a fermionic superfield. Since comparatively little is known concerning such extensions, we propose to construct an $N=2$ supersymmetric extension of the system (\ref{b16}) in which the superfield is bosonic. This is motivated by the fact that supersymmetric versions of the Schr\"{o}dinger equation have recently been considered \cite{Unterberger}. The question arises as the whether a link could be determined between the supersymmetric versions of the Schr\"{o}dinger and hydrodynamic equations and their subalgebras which is similar to the link established for the classical versions of the Schr\"{o}dinger and hydrodynamic systems. In this paper, we concentrate on the first step: building a supersymmetric extension of the hydrodynamic system (\ref{b16}) using a bosonic superfield and determining its symmetry properties. This cannot be done in a non-trivial way by including only one fermionic independent variable ($N=1$), so we will extend the space of independent variables to a superspace involving two fermionic independent variable ($N=2$). In addition, we provide a number of interesting exact, analytic solutions of our extended system. These include polynomial, radical and exponential solutions, bumps, kinks and multisolitons as well as solutions expressed in terms of quadratures.

This paper is organized as follows. In Section 2, we construct an $N=2$ supersymmetric extension of the system (\ref{b16}) through the use of a bosonic superfield involving two independent fermionic variables. Section 3 involves a computation of a Lie superalgebra of infinitesimal symmetries of our extended hydrodynamic system. In Section 4, we perform a number of symmetry reductions and present nine invariant solutions of the supersymmetric hydrodynamic model. Finally, Section 5 contains concluding remarks and a description of possible future developments. In Appendix A we present the full list of 401 one-dimensional Lie subalgebras associated with the supersymmetric system.

\section{$N=2$ supersymmetric version of the system in Riemann invariants}

In order to construct an $N=2$ supersymmetric extension of the hydrodynamic type system (\ref{b16}), we enlarge the 
space of independent variables $\{x,t\}$ to a superspace $\{x,t,\theta,\varphi\}$. Here, $x$ and $t$ are the standard bosonic space and time variables for $(1+1)$-dimensional space, while $\theta$ and $\varphi$ are independent fermionic Grassmann variables. In addition, we replace the two bosonic fields $R(x,t)$ and $S(x,t)$ by the bosonic superfields
\begin{equation}
{\mathcal A}(x,t,\theta,\varphi)=U(x,t)+\theta \eta(x,t)+\varphi \pi(x,t)+\theta\varphi R(x,t),
\end{equation}
and
\begin{equation}
{\mathcal B}(x,t,\theta,\varphi)=V(x,t)+\theta \psi(x,t)+\varphi \omega(x,t)+\theta\varphi S(x,t),
\label{e2}
\end{equation}
where $\eta(x,t)$, $\psi(x,t)$, $\pi(x,t)$ and $\omega(x,t)$ are four new fermionic-valued fields, and $U(x,t)$ and $V(x,t)$ are two additional bosonic fields. We construct our generalization in such a way that it is invariant under the supersymmetry transformations
\begin{equation}
x\rightarrow x+\underline{\alpha}\theta,\quad \theta\rightarrow\theta-\underline{\alpha},
\label{e3}
\end{equation}
and
\begin{equation}
t\rightarrow t+\underline{\beta}\varphi,\quad \varphi\rightarrow\varphi-\underline{\beta},
\label{e3A}
\end{equation}
where $\underline{\alpha}$ and $\underline{\beta}$ are constant fermionic parameters. The transformations (\ref{e3}) and (\ref{e3A}) are generated by the infinitesimal supersymmetry generators
\begin{equation}
Q_x=\theta\partial_x-\partial_{\theta}\qquad\mbox{ and }\qquad Q_t=\varphi\partial_t-\partial_{\varphi},
\label{e4}
\end{equation}
respectively, where $\partial_x=\partial/\partial x$, etc. In order to make our superfield theory manifestly invariant under the action of $Q_x$ and $Q_t$, it is useful to introduce the covariant derivatives
\begin{equation}
D_x=\theta\partial_x+\partial_{\theta}\qquad\mbox{ and }\qquad D_t=\varphi\partial_t+\partial_{\varphi},
\label{e4bis}
\end{equation}
which possess the property that they anticommute with the operators $Q_x$ and $Q_t$, i.e.
\begin{equation}
\{Q_x,D_x\}=\{Q_x,D_t\}=\{Q_t,D_x\}=\{Q_t,D_t\}=0.
\label{anticommm}
\end{equation}
The most general form of a supersymmetric extension of system (\ref{b16}), expressed in terms of superfields ${\mathcal A}$ and ${\mathcal B}$ and covariant derivatives $D_x$ and $D_t$, is written in the form
\begin{equation}
\begin{split}
D_t^2{\mathcal A}&+a_1{\mathcal B}\left(D_x^3D_t{\mathcal A}\right)+a_2\left(D_x{\mathcal B}\right)\left(D_x^2D_t{\mathcal A}\right)+a_3\left(D_t{\mathcal B}\right)\left(D_x^3{\mathcal A}\right)\\ &+(a_2-a_1-a_3-1)\left(D_xD_t{\mathcal B}\right)\left(D_x^2{\mathcal A}\right)=0,\\
\end{split}
\label{e5m}
\end{equation}
\begin{equation}
\begin{split}
D_t^2{\mathcal B}&+a_4{\mathcal A}\left(D_x^3D_t{\mathcal B}\right)+a_5\left(D_x{\mathcal A}\right)\left(D_x^2D_t{\mathcal B}\right)+a_6\left(D_t{\mathcal A}\right)\left(D_x^3{\mathcal B}\right)\\ &+(a_5-a_4-a_6-1)\left(D_xD_t{\mathcal A}\right)\left(D_x^2{\mathcal B}\right)=0,
\end{split}
\label{e5}
\end{equation}
where $a_i$, $i=1,\ldots,6$ are arbitrary real constant parameters. The supersymmetric system formed by equations (\ref{e5m}) and (\ref{e5}) can be decomposed into the following eight differential equations corresponding to the coefficients of the various powers of $\theta$ and $\varphi$
\begin{equation}
\begin{split}
R_t&+SR_x+(a_1-a_2)\psi\eta_{xt}+(a_1-a_2+1)\psi_t\eta_x+(a_1+a_3)\omega\pi_{xx}\\ &+(a_1+a_3+1)\omega_x\pi_x+a_1VU_{xxt}+a_2V_xU_{xt}-a_3V_tU_{xx}\\ &+(a_2-a_1-a_3-1)U_xV_{xt}=0,\\ & \\
S_t&+RS_x+(a_4-a_5)\eta\psi_{xt}+(a_4-a_5+1)\eta_t\psi_x+(a_4+a_6)\pi\omega_{xx}\\&+(a_4+a_6+1)\pi_x\omega_x +a_4UV_{xxt}+a_5U_xV_{xt}-a_6U_tV_{xx}\\&+(a_5-a_4-a_6-1)V_xU_{xt}=0,\\ & \\
\eta_t&+(a_2-a_1)\psi R_x+a_1V\pi_{xx}+a_2V_x\pi_x+(a_1-a_2+1)S\eta_x\\ &-a_3\omega U_{xx}+(a_2-a_1-a_3-1)U_x\omega_x=0,\\ & \\
\psi_t&+(a_5-a_4)\eta S_x+a_4U\omega_{xx}+a_5U_x\omega_x+(a_4-a_5+1)R\psi_x\\ &-a_6\pi V_{xx}+(a_5-a_4-a_6-1)V_x\pi_x=0,\\ & \\
\pi_t&-(a_1+a_3)\omega R_x-a_1V\eta_{xt}+a_3V_t\eta_x+(a_1+a_3+1)S\pi_x\\ &-a_2\psi U_{xt}+(a_1-a_2+a_3+1)\psi_tU_x=0,\\ & \\
\omega_t&-(a_4+a_6)\pi S_x-a_4U\psi_{xt}+a_6U_t\psi_x+(a_4+a_6+1)R\omega_x\\ &-a_5\eta V_{xt}+(a_4-a_5+a_6+1)\eta_tV_x=0,\\ & \\
U_t&-a_1VR_x+a_2\psi\pi_x+a_3\omega\eta_x+(a_1-a_2+a_3+1)SU_x=0,\\ & \\
V_t&-a_4US_x+a_5\eta\omega_x+a_6\pi\psi_x+(a_4-a_5+a_6+1)RV_x=0.
\end{split}
\label{susyeqq8}
\end{equation}
In this paper, we consider the case where the real constant parameters vanish (i.e. $a_i=0$, $i=1,\ldots 6$). 
Here, the supersymmetric hydrodynamic system described by equations (\ref{e5m}) and (\ref{e5}) becomes
\begin{equation}
\begin{split}
& D_t^2{\mathcal A}-\left(D_xD_t{\mathcal B}\right)\left(D_x^2{\mathcal A}\right)=0,\qquad D_t^2{\mathcal B}-\left(D_xD_t{\mathcal A}\right)\left(D_x^2{\mathcal B}\right)=0,
\end{split}
\label{e7}
\end{equation}
and the system (\ref{susyeqq8}) takes the form
\begin{equation}
\begin{split}
& R_t+SR_x+\psi_t\eta_x+\omega_x\pi_x-U_xV_{xt}=0,\qquad
 S_t+RS_x+\eta_t\psi_x+\pi_x\omega_x-V_xU_{xt}=0,\\
& \eta_t+S\eta_x-U_x\omega_x=0,\qquad
 \psi_t+R\psi_x-V_x\pi_x=0,\qquad
 \pi_t+S\pi_x+\psi_tU_x=0,\\
& \omega_t+R\omega_x+\eta_tV_x=0,\qquad
 U_t+SU_x=0,\qquad
 V_t+RV_x=0.
\end{split}
\label{e8}
\end{equation}
It should be noted that the system (\ref{e8}) is invariant under the discrete transformation $R\leftrightarrow S,\eta\leftrightarrow\psi,\pi\leftrightarrow \omega,U\leftrightarrow V$.
In what follows, we will refer to the system (\ref{e8}) as the $N=2$ supersymmetric extension of the system (\ref{b16}) in Riemann invariants. In the limiting case where the four fermionic fields $\eta$, $\psi$, $\pi$, $\omega$ and the two additional bosonic fields $U$ and $V$ tend to zero, we recover the classical version of the hydrodynamic system (\ref{b16}).

\section{Symmetries of the supersymmetric hydrodynamic model}

\subsection{Lie point symmetries}

Using the techniques described in \cite{Olver} adapted to Grassmann-valued differential equations, we determine a Lie superalgebra ${\mathcal L}$ of infinitesimal symmetries of the supersymmetric system (\ref{e8}). This superalgebra is generated by the following twelve fiber-preserving vector fields
\begin{equation}
\begin{split}
& P_0=\partial_t,\qquad P_1=\partial_x,\qquad T_1=\partial_U,\qquad T_2=\partial_V,\qquad Z_1=\partial_{\eta},\qquad Z_2=\partial_{\psi},\\ 
& Z_3=\partial_{\pi},\quad Z_4=\partial_{\omega},\quad M_1=x\partial_x+R\partial_R+S\partial_S+2\psi\partial_{\psi}+3\omega\partial_{\omega}-U\partial_U+4V\partial_V,\\ & M_2=t\partial_t-R\partial_R-S\partial_S-\psi\partial_{\psi}-2\omega\partial_{\omega}+U\partial_U-2V\partial_V,\\
& M_3=\eta\partial_{\eta}-\psi\partial_{\psi}+U\partial_U-V\partial_V,\qquad M_4=\pi\partial_{\pi}-\omega\partial_{\omega}+U\partial_U-V\partial_V.
\end{split}
\label{f1}
\end{equation}
The physical interpretation of this Lie superalgebra as it applies to the coordinates $(x,t,R,S,\eta,\psi,\pi,\omega,U,V)$ is as follows. The vector fields $P_0$, $P_1$, $T_1$, $T_2$, $Z_1$, $Z_2$, $Z_3$ and $Z_4$ generate translations in $x$, $t$, $U$, $V$, $\eta$, $\psi$, $\pi$ and $\omega$ respectively, while $M_1$, $M_2$, $M_3$ and $M_4$ correspond to four independent dilations involving the independent and dependent (bosonic and fermionic) variables. The supercommutation  relations of the generators described in (\ref{f1}) are summarized in Table 1.

\begin{table}[htbp]
  \begin{center}
\caption{Supercommutation table for the Lie superalgebra ${\cal L}$ spanned by the
  vector fields (\ref{f1})}
\vspace{5mm}
\setlength{\extrarowheight}{4pt}
\begin{tabular}{|c||c|c|c|c|c|c|c|c|c|c|c|c|}\hline
& $\mathbf{M_1}$ & $\mathbf{M_2}$ & $\mathbf{M_3}$ & $\mathbf{M_4}$ & $\mathbf{P_0}$ & $\mathbf{P_1}$ & $\mathbf{T_1}$ & $\mathbf{T_2}$ & $\mathbf{Z_1}$ & $\mathbf{Z_2}$ & $\mathbf{Z_3}$ & $\mathbf{Z_4}$\\[0.5ex]\hline\hline
$\mathbf{M_1}$ & $0$ & $0$ & $0$ & $0$ & $0$ & $-P_1$ & $T_1$ & $-4T_2$ & $0$ & $-2Z_2$ & $0$ & $-3Z_4$\\\hline
$\mathbf{M_2}$ & $0$ & $0$ & $0$ & $0$ & $-P_0$ & $0$ & $-T_1$ & $2T_2$ & $0$ & $Z_2$ & $0$ & $2Z_4$\\\hline
$\mathbf{M_3}$ & $0$ & $0$ & $0$ & $0$ & $0$ & $0$ & $-T_1$ & $T_2$ & $-Z_1$ & $Z_2$ & $0$ & $0$\\\hline
$\mathbf{M_4}$ & $0$ & $0$ & $0$ & $0$ & $0$ & $0$ & $-T_1$ & $T_2$ & $0$ & $0$ & $-Z_3$ & $Z_4$\\\hline
$\mathbf{P_0}$ & $0$ & $P_0$ & $0$ & $0$ & $0$ & $0$ & $0$ & $0$ & $0$ & $0$ & $0$ & $0$\\\hline
$\mathbf{P_1}$ & $P_1$ & $0$ & $0$ & $0$ & $0$ & $0$ & $0$ & $0$ & $0$ & $0$ & $0$ & $0$\\\hline
$\mathbf{T_1}$ & $-T_1$ & $T_1$ & $T_1$ & $T_1$ & $0$ & $0$ & $0$ & $0$ & $0$ & $0$ & $0$ & $0$\\\hline
$\mathbf{T_2}$ & $4T_2$ & $-2T_2$ & $-T_2$ & $-T_2$ & $0$ & $0$ & $0$ & $0$ & $0$ & $0$ & $0$ & $0$\\\hline
$\mathbf{Z_1}$ & $0$ & $0$ & $Z_1$ & $0$ & $0$ & $0$ & $0$ & $0$ & $0$ & $0$ & $0$ & $0$\\\hline
$\mathbf{Z_2}$ & $2Z_2$ & $-Z_2$ & $-Z_2$ & $0$ & $0$ & $0$ & $0$ & $0$ & $0$ & $0$ & $0$ & $0$\\\hline
$\mathbf{Z_3}$ & $0$ & $0$ & $0$ & $Z_3$ & $0$ & $0$ & $0$ & $0$ & $0$ & $0$ & $0$ & $0$\\\hline
$\mathbf{Z_4}$ & $3Z_4$ & $-2Z_4$ & $0$ & $-Z_4$ & $0$ & $0$ & $0$ & $0$ & $0$ & $0$ & $0$ & $0$\\\hline
\end{tabular}
  \end{center}
\end{table}

\subsection{Subalgebras of the Lie superalgebra}

We now wish to classify the one-dimensional subalgebras of the superalgebra ${\mathcal L}$. That is, we want to construct a list of representative subalgebras of ${\mathcal L}$ such that each subalgebra of ${\mathcal L}$ is conjugate to one and only one element of the list under the Baker-Campbell-Hausdorff equivalence conjugation relation
\begin{equation}
X\rightarrow e^YXe^{-Y}=X+[Y,X]+{1\over 2!}\left[Y,[Y,X]\right]+{1\over 3!}\left[Y,\left[Y,[Y,X]\right]\right]+\ldots
\label{f3}
\end{equation}
The classification methods do not readily lend themselves to computerization and therefore require a great deal of tedious, very involved computation. The full list of one-dimensional subalgebras of the superalgebra ${\mathcal L}$ is quite voluminous (a total of 401 subalgebras), and so we provide only a brief discussion in this section. Complete details of this list are provided in Appendix A.

In order to perform the classification, we first decompose ${\mathcal L}$ into the following composite semidirect sum
\begin{equation}
\begin{split}
{\cal L}= \Bigg{\{}\Bigg{\{}\bigg{\{}\bigg{\{}\Big{\{}\Big{\{}&\big{\{}{\mathcal M}^{(0)}\sdir\{P_0\}\big{\}}\sdir\{P_1\}\Big{\}}\sdir\{T_1\}\Big{\}}\sdir\{T_2\}\bigg{\}}\\ & \sdir\{Z_1\}\bigg{\}}\sdir\{Z_2\}\Bigg{\}}\sdir\{Z_3\}\Bigg{\}}\sdir\{Z_4\},
\end{split}
\label{f2}
\end{equation}
where we introduce the notation ${\mathcal M}^{(0)}=\{M_1,M_2,M_3,M_4\}$.
The one-dimensional subalgebras of ${\cal L}$ are classified using techniques for semidirect sums of algebras as described in \cite{Winternitz}. In general, the representative subalgebras of a semidirect sum ${\mathcal F}\sdir {\mathcal N}$ of two Lie algebras ${\mathcal F}$ and ${\mathcal N}$ can be categorized as splitting subalgebras (which can be written in the form ${\mathcal F}_0\sdir {\mathcal N}_0$ where ${\mathcal F}_0\subset{\mathcal F}$ and ${\mathcal N}_0\subset{\mathcal N}$) and nonsplitting subalgebras (all representative subalgebras which are not conjugate to a splitting one).\\
For each step of the form ${\mathcal M}\sdir {\mathcal P}$ in the composite semidirect sum, we calculate the splitting and nonsplitting subalgebras. We obtain subalgebras of the form
\begin{equation}
\{b_1M_1+b_2M_2+b_3M_3+b_4M_4+c_1P_0+c_2P_1+c_3T_1+c_4T_2+\underline{\alpha}Z_1+\underline{\beta}Z_2+\underline{\gamma}Z_3+\underline{\delta}Z_4\},
\end{equation}
where $b_1$, $b_2$, $b_3$ and $b_4$ are real bosonic constants, $c_1$, $c_2$, $c_3$ and $c_4$ generally represent the values $1$, $-1$ or $0$, while $\underline{\alpha}$, $\underline{\beta}$, $\underline{\gamma}$ and $\underline{\delta}$ are fermionic constants. Not all arbitrary values of the parameters will result in separate classes of subalgebras since the subalgebras corresponding to many distinct combinations of parameters can be linked through the action of the Lie group generated by ${\mathcal L}$.

\section{Invariant Solutions}

Since the list of one-dimensional subalgebra classes of the Lie superalgebra ${\cal L}$ is so vast, we will not attempt to make a comprehensive analysis of all symmetry reductions and group-invariant solutions of the supersymmetric hydrodynamic system (\ref{e8}). Instead, we select subalgebras from the list and use the symmetry reduction method in order to obtain certain interesting examples of solutions of (\ref{e8}) which are invariant with respect to each respective subalgebra. Specifically, we choose
\begin{equation}
\begin{split}
& {\mathcal L}_4=\{M_1\}\mbox{ }(\mbox{where }a=b=c=0),\qquad {\mathcal L}_5=\{P_0\},\\ & {\mathcal L}_7=\{M_3+\varepsilon P_0\}\mbox{ }(\mbox{where }a=0),\qquad {\mathcal L}_{10}=\{M_4+\varepsilon P_1\},\\ & {\mathcal L}_{13}=\{M_4+\varepsilon P_0+\mu P_1\},\qquad\qquad {\mathcal L}_{15}=\{P_0+\varepsilon P_1\},\\ & {\mathcal L}_{68}=\{P_0+\varepsilon P_1+\mu T_1+\underline{\alpha}Z_1\},\qquad {\mathcal L}_{149}=\{P_0+\varepsilon P_1+\underline{\alpha}Z_3\}. 
\end{split}
\label{selected}
\end{equation}
For each of the subalgebras mentioned in (\ref{selected}), we begin by finding the nine invariants associated to the subalgebras along with the corresponding change of variable. The symmetry variable is labelled with the symbol $\xi$. Next, by substituting the change of variable and its derivatives into the original system (\ref{e8}), we obtain a system of reduced ordinary differential equations. These results are listed in Tables 2 and 3.

\begin{table}[htbp]
  \begin{center}
\caption{Invariants and change of variable for selected subalgebras of the Lie superalgebra ${\cal L}$ spanned by the
  vector fields (\ref{f1})}
\setlength{\extrarowheight}{4pt}
\begin{tabular}{|c|c|c|}\hline
Subalgebra & Invariants & Relations and
Change of Variable\\[0.5ex]\hline\hline
${\mathcal L}_4=\{M_1\}$ & $\xi={1\over t}$, ${1\over x}R$, ${1\over x}S$, $\eta$,  & $R=xF(\xi)$, $S=xG(\xi)$, $\eta=\eta(\xi)$,  \\
  & ${1\over x^2}\psi$, $\pi$, ${1\over x^3}\omega$, $xU$, ${1\over x^4}V$ & $\psi=x^2\Psi(\xi)$, $\pi=\pi(\xi)$, $\omega=x^3\Omega(\xi)$,  \\
  &  & $U={1\over x}{\mathcal Y}(\xi)$, $V=x^4{\mathcal Z}(\xi)$ \\\hline
${\mathcal L}_5=\{P_0\}$ & $x$, $R$, $S$, $\eta$, $\psi$, & $R=R(x)$, $S=S(x)$, $\eta=\eta(x)$, \\
  & $\pi$, $\omega$, $U$, $V$ & $\psi=\psi(x)$, $\pi=\pi(x)$, $\omega=\omega(x)$,  \\
  &  & $U=U(x)$, $V=V(x)$ \\\hline
${\mathcal L}_7=\{M_3+\varepsilon P_0\}$ & $x$, $R$, $S$, $e^{-\varepsilon t}\eta$,  & $R=R(x)$, $S=S(x)$, $\eta=e^{\varepsilon t}{\mathcal H}(x)$,  \\
& $e^{\varepsilon t}\psi$, $\pi$, $\omega$, $e^{-\varepsilon t}U$, $e^{\varepsilon t}V$ & $\psi=e^{-\varepsilon t}\Psi(x)$, $\pi=\pi(x)$, $\omega=\omega(x)$,  \\
  &  & $U=e^{\varepsilon t}{\mathcal Y}(x)$, $V=e^{-\varepsilon t}{\mathcal Z}(x)$ \\\hline
${\mathcal L}_{10}=\{M_4+\varepsilon P_1\}$ & $t$, $R$, $S$, $\eta$,  & $R=R(t)$, $S=S(t)$, $\eta=\eta(t)$,  \\
& $\psi$, $e^{-\varepsilon x}\pi$, $e^{\varepsilon x}\omega$, $e^{-\varepsilon x}U$, $e^{\varepsilon x}V$ & $\psi=\psi(t)$, $\pi=e^{\varepsilon x}{\mathcal P}(t)$, $\omega=e^{-\varepsilon x}\Omega(\xi)$,  \\
  &  & $U=e^{\varepsilon x}{\mathcal Y}(t)$, $V=e^{-\varepsilon x}{\mathcal Z}(t)$ \\\hline
${\mathcal L}_{13}=\{M_4+\varepsilon P_0+\mu P_1\}$ & $\xi=x-\varepsilon\mu t$, $R$, $S$, $\eta$,  & $R=R(\xi)$, $S=S(\xi)$, $\eta=\eta(\xi)$,  \\
& $\psi$, $e^{-\varepsilon t}\pi$, $e^{\varepsilon t}\omega$, $e^{-\varepsilon t}U$, $e^{\varepsilon t}V$ & $\psi=\psi(\xi)$, $\pi=e^{\varepsilon t}{\mathcal P}(\xi)$, $\omega=e^{-\varepsilon t}\Omega(\xi)$,  \\
  &  & $U=e^{\varepsilon t}{\mathcal Y}(\xi)$, $V=e^{-\varepsilon t}{\mathcal Z}(\xi)$ \\\hline
${\mathcal L}_{15}=\{P_0+\varepsilon P_1\}$ & $\xi=x-\varepsilon t$, $R$, $S$, $\eta$,  & $R=R(\xi)$, $S=S(\xi)$, $\eta=\eta(\xi)$,  \\
  & $\psi$, $\pi$, $\omega$, $U$, $V$ & $\psi=\psi(\xi)$, $\pi=\pi(\xi)$, $\omega=\omega(\xi)$,  \\
  &  & $U=U(\xi)$, $V=V(\xi)$ \\\hline
${\mathcal L}_{68}=\{P_0+\varepsilon P_1+\mu T_1+\underline{\alpha}Z_1\}$ & $\xi=x-\varepsilon t$, $R$, $S$, $\eta-\underline{\alpha}t$,  & $R=R(\xi)$, $S=S(\xi)$, $\eta={\mathcal H}(\xi)+\underline{\alpha}t$,  \\
  & $\psi$, $\pi$, $\omega$, $U-\mu t$, $V$ & $\psi=\psi(\xi)$, $\pi=\pi(\xi)$, $\omega=\omega(\xi)$,  \\
  &  & $U={\mathcal Y}(\xi)+\mu t$, $V=V(\xi)$ \\\hline
${\mathcal L}_{149}=\{P_0+\varepsilon P_1+\underline{\alpha}Z_3\}$ & $\xi=x-\varepsilon t$, $R$, $S$, $\eta$,  & $R=R(\xi)$, $S=S(\xi)$, $\eta=\eta(\xi)$,  \\
  & $\psi$, $\pi-\underline{\alpha}t$, $\omega$, $U$, $V$ & $\psi=\psi(\xi)$, $\pi={\mathcal P}(\xi)+\underline{\alpha}t$, $\omega=\omega(\xi)$,  \\
  &  & $U=U(\xi)$, $V=V(\xi)$ \\\hline
\end{tabular}
  \end{center}
\end{table}

\begin{table}[htbp]
  \begin{center}
\caption{Reduced Equations obtained for selected subalgebras of the Lie superalgebra ${\cal L}$ spanned by the
  vector fields (\ref{f1})}
\vspace{5mm}
\setlength{\extrarowheight}{4pt}
\begin{tabular}{|c|c|}\hline
Subalgebra & Reduced Equations \\[0.5ex]\hline\hline
${\mathcal L}_4=\{M_1\}$ & $-\xi^2F_{\xi}+FG-4\xi^2{\mathcal Y}{\mathcal Z}_{\xi}=0$,\quad $-\xi^2G_{\xi}+FG-2\xi^2{\mathcal H}_{\xi}\Psi-4\xi^2{\mathcal Z}{\mathcal Y}_{\xi}=0$, \\ 
& $-\xi^2{\mathcal H}_{\xi}+3{\mathcal Y}\Omega=0$,\quad $-\xi^2\Psi_{\xi}+2F\Psi=0$,\quad $-\xi^2{\mathcal P}_{\xi}+\xi^2{\mathcal Y}\Psi_{\xi}=0$,\\
& $-\xi^2\Omega_{\xi}+3F\Omega-4\xi^2{\mathcal Z}{\mathcal H}_{\xi}=0$,\qquad $-\xi^2{\mathcal Y}_{\xi}-G{\mathcal Y}=0$, \\
& $-\xi^2{\mathcal Z}_{\xi}+4F{\mathcal Z}=0$ \\\hline
${\mathcal L}_5=\{P_0\}$ & $SR_x+\omega_x\pi_x=0$,\qquad $RS_x+\pi_x\omega_x=0$,\qquad $S\eta_x-U_x\omega_x=0$, \\ 
& $R\psi_x-V_x\pi_x=0$,\quad $S\pi_x=0$,\quad $R\omega_x=0$,\quad $SU_x=0$,\quad $RV_x=0$\\\hline
${\mathcal L}_7=\{M_3+\varepsilon P_0\}$ & $SR_x-\varepsilon\Psi{\mathcal H}_x+\omega_x\pi_x+\varepsilon{\mathcal Y}_x{\mathcal Z}_x=0$,\\ 
& $RS_x+\varepsilon{\mathcal H}\Psi_x+\pi_x\omega_x-\varepsilon{\mathcal Y}_x{\mathcal Z}_x=0$, \\
& $\varepsilon{\mathcal H}+S{\mathcal H}_x-{\mathcal Y}_x\omega_x=0$,\qquad $-\varepsilon\Psi+R\Psi_x-{\mathcal Z}_x\pi_x=0$, \\
& $S\pi_x-\varepsilon{\mathcal Y}_x\Psi=0$,\qquad $R\omega_x+\varepsilon{\mathcal H}{\mathcal Z}_x=0$, \\
& $\varepsilon{\mathcal Y}+S{\mathcal Y}_x=0$,\qquad $-\varepsilon{\mathcal Z}+R{\mathcal Z}_x=0$ \\\hline
${\mathcal L}_{10}=\{M_4+\varepsilon P_1\}$ & $R_t-\Omega{\mathcal P}+{\mathcal Y}{\mathcal Z}_t=0$,\qquad $S_t-{\mathcal P}\Omega+{\mathcal Z}{\mathcal Y}_t=0$, \\ 
& $\eta_t+{\mathcal Y}\Omega=0$,\qquad $\psi_t+{\mathcal Z}{\mathcal P}=0$,\\
& ${\mathcal P}_t+\varepsilon S{\mathcal P}+\varepsilon{\mathcal Y}\psi_t=0$,\qquad $\Omega_t-\varepsilon R\Omega-\varepsilon{\mathcal Z}\eta_t=0$, \\
& ${\mathcal Y}_t+\varepsilon S{\mathcal Y}=0$,\qquad ${\mathcal Z}_t-\varepsilon R{\mathcal Z}=0$ \\\hline
${\mathcal L}_{13}=\{M_4+\varepsilon P_0+\mu P_1\}$ & $(S-\varepsilon\mu)R_{\xi}-\varepsilon\mu\psi_{\xi}\eta_{\xi}+\Omega_{\xi}{\mathcal P}_{\xi}+\varepsilon{\mathcal Y}_{\xi}{\mathcal Z}_{\xi}+\varepsilon\mu{\mathcal Y}_{\xi}{\mathcal Z}_{\xi\xi}=0$,\\ 
& $(R-\varepsilon\mu)S_{\xi}-\varepsilon\mu\eta_{\xi}\psi_{\xi}+{\mathcal P}_{\xi}\Omega_{\xi}-\varepsilon{\mathcal Y}_{\xi}{\mathcal Z}_{\xi}+\varepsilon\mu{\mathcal Z}_{\xi}{\mathcal Y}_{\xi\xi}=0$, \\
& $(S-\varepsilon\mu)\eta_{\xi}-{\mathcal Y}_{\xi}\Omega_{\xi}=0$,\qquad $(R-\varepsilon\mu)\psi_{\xi}-{\mathcal Z}_{\xi}{\mathcal P}_{\xi}=0$, \\
& $(S-\varepsilon\mu){\mathcal P}_{\xi}+\varepsilon{\mathcal P}-\varepsilon\mu{\mathcal Y}_{\xi}\psi_{\xi}=0$,\quad $(R-\varepsilon\mu)\Omega_{\xi}-\varepsilon\Omega-\varepsilon\mu\eta_{\xi}{\mathcal Z}_{\xi}=0$, \\
& $(S-\varepsilon\mu){\mathcal Y}_{\xi}+\varepsilon{\mathcal Y}=0$,\qquad $(R-\varepsilon\mu){\mathcal Z}_{\xi}-\varepsilon{\mathcal Z}=0$ \\\hline
${\mathcal L}_{15}=\{P_0+\varepsilon P_1\}$ & $(S-\varepsilon)R_{\xi}-\varepsilon\psi_{\xi}\eta_{\xi}+\omega_{\xi}\pi_{\xi}+\varepsilon U_{\xi}V_{\xi\xi}=0$,\\
& $(R-\varepsilon)S_{\xi}-\varepsilon\eta_{\xi}\psi_{\xi}+\pi_{\xi}\omega_{\xi}+\varepsilon V_{\xi}U_{\xi\xi}=0$, \\ 
& $(S-\varepsilon)\eta_{\xi}-U_{\xi}\omega_{\xi}=0$,\qquad  $(R-\varepsilon)\psi_{\xi}-V_{\xi}\pi_{\xi}=0$,\\
& $(S-\varepsilon)\pi_{\xi}-\varepsilon U_{\xi}\psi_{\xi}=0$,\qquad  $(R-\varepsilon)\omega_{\xi}-\varepsilon\eta_{\xi}V_{\xi}=0$, \\
& $(S-\varepsilon)U_{\xi}=0$,\qquad $(R-\varepsilon)V_{\xi}=0$ \\\hline
${\mathcal L}_{68}=\{P_0+\varepsilon P_1+\mu T_1+\underline{\alpha}Z_1\}$ & $(S-\varepsilon)R_{\xi}-\varepsilon\psi_{\xi}{\mathcal H}_{\xi}+\omega_{\xi}\pi_{\xi}+\varepsilon{\mathcal Y}_{\xi}V_{\xi\xi}=0$,\\ 
& $(R-\varepsilon)S_{\xi}+\underline{\alpha}\psi_{\xi}-\varepsilon{\mathcal H}_{\xi}\psi_{\xi}+\pi_{\xi}\omega_{\xi}+\varepsilon V_{\xi}{\mathcal Y}_{\xi\xi}=0$, \\
& $(S-\varepsilon){\mathcal H}_{\xi}+\underline{\alpha}-{\mathcal Y}_{\xi}\omega_{\xi}=0$,\qquad $(R-\varepsilon)\psi_{\xi}-V_{\xi}\pi_{\xi}=0$, \\
& $(S-\varepsilon)\pi_{\xi}-\varepsilon{\mathcal Y}_{\xi}\psi_{\xi}=0$,\qquad $(R-\varepsilon)\omega_{\xi}+\underline{\alpha}V_{\xi}-\varepsilon{\mathcal H}_{\xi}V_{\xi}=0$, \\
& $(S-\varepsilon){\mathcal Y}_{\xi}+\mu=0$,\qquad $(R-\varepsilon)V_{\xi}=0$ \\\hline
${\mathcal L}_{149}=\{P_0+\varepsilon P_1+\underline{\alpha}Z_3\}$ & $(S-\varepsilon)R_{\xi}-\varepsilon\psi_{\xi}\eta_{\xi}+\omega_{\xi}{\mathcal P}_{\xi}+\varepsilon U_{\xi}V_{\xi\xi}=0$,\\
& $(R-\varepsilon)S_{\xi}-\varepsilon\eta_{\xi}\psi_{\xi}+{\mathcal P}_{\xi}\omega_{\xi}+\varepsilon V_{\xi}U_{\xi\xi}=0$, \\ 
& $(S-\varepsilon)\eta_{\xi}-U_{\xi}\omega_{\xi}=0$,\qquad  $(R-\varepsilon)\psi_{\xi}-V_{\xi}{\mathcal P}_{\xi}=0$,\\
& $\underline{\alpha}+(S-\varepsilon){\mathcal P}_{\xi}-\varepsilon U_{\xi}\psi_{\xi}=0$,\qquad  $(R-\varepsilon)\omega_{\xi}-\varepsilon\eta_{\xi}V_{\xi}=0$, \\
& $(S-\varepsilon)U_{\xi}=0$,\qquad $(R-\varepsilon)V_{\xi}=0$ \\\hline
\end{tabular}
  \end{center}
\end{table}

From certain specific solutions of the reduced system, we use the change of variable in reverse in order to obtain an invariant solution of the supersymmetric hydrodynamic system (\ref{e8}). In what follows, $U_0$, $V_0$, $n$, $k$, $A$, $B$, $C_1$ and $C_2$ represent arbitrary bosonic constants, while $\underline{\eta_0}$, $\underline{\pi_0}$, $\underline{\omega_0}$, $\underline{D_1}$, $\underline{D_2}$, $\underline{D_3}$, $\underline{E_1}$, $\underline{E_2}$, $\underline{F_1}$ and $\underline{F_2}$ represent arbitrary fermionic constants.

We begin by considering subalgebra ${\mathcal L}_4$, where we obtain the following solution
\begin{equation}
\begin{split}
& R=-{kx\over 1+C_1ke^{kt}},\qquad S={C_1k^2xe^{kt}\over 1+C_1ke^{kt}},\qquad \eta=\underline{D_2},\\ & \psi=\underline{D_3}x^2\exp{\int{-2k\mbox{d}\xi\over \xi^2(1+C_1ke^{k/\xi})}\bigg{|}_{\xi=1/t}},\qquad \pi=\underline{D_1},\\ & \omega=\underline{D_4}x^3\exp{\int{-3k\mbox{d}\xi\over \xi^2(1+C_1ke^{k/\xi})}\bigg{|}_{\xi=1/t}},\qquad U=0,\\ & V=C_2x^4\exp{\int{-4k\mbox{d}\xi\over \xi^2(1+C_1ke^{k/\xi})}\bigg{|}_{\xi=1/t}}.
\end{split}
\label{solution2}
\end{equation}
Here, the classical bosonic fields $R$ and $S$ are kink solutions, while the fermionic fields $\psi$ and $\omega$ and the bosonic field $V$ are expressed in terms of quadratures.

An invariant solution corresponding to subalgebra ${\mathcal L}_5$ is
\begin{equation}
\begin{split}
& R=\underline{D_1}\underline{E_1}Ax^n,\qquad S=\underline{D_2}\underline{E_2}Bx^{-n},\qquad \eta=\underline{E_2}f_2(x),\qquad \psi=\underline{E_1}f_1(x),\\
& \pi=2\underline{D_2}\underline{E_2}\underline{F_2}Bx^{1/2},\qquad \omega=2\underline{D_1}\underline{E_1}\underline{F_1}Ax^{1/2},\qquad U=U_0,\qquad V=V_0,
\end{split}
\label{solution3}
\end{equation}
where $\underline{F_1}\underline{F_2}=-n$, and $f_1$ and $f_2$ are arbitrary bosonic functions of $x$. This is a static solution, where $r$, $S$, $\pi$ and $\omega$ are expressed in terms of radicals, while $\eta$ and $\psi$ are fermionic solutions of arbitrary shape. Certain special functions, such as bumps, kinks and multiple waves can be used to model static physical phenomena.

For subalgebra ${\mathcal L}_7$, an invariant solution is represented by
\begin{equation}
\begin{split}
& R=\varepsilon x,\qquad S=-\varepsilon x,\qquad \eta=\underline{D_1}xe^{\varepsilon t},\qquad \psi=\underline{D_2}xe^{-\varepsilon t},\\
& \pi=\underline{\pi_0},\qquad \omega=-C_1\underline{D_1}x+\underline{D_3},\qquad U=0,\qquad V=C_1xe^{-\varepsilon t},
\end{split}
\label{solution5}
\end{equation}
where $\underline{D_1}\underline{D_2}=\varepsilon$,
which is exponential in $t$, but linear in $x$.

Next, we have the following two exponential solutions involving travelling waves.

Subalgebra ${\mathcal L}_{10}$ leads to the solution
\begin{equation}
\begin{split}
& R=\varepsilon,\qquad S=\varepsilon,\qquad \eta={\varepsilon\over C_1}\underline{D_2}e^{-\varepsilon C_1C_2t},\qquad \psi=-{\varepsilon\over C_2}\underline{D_1}e^{\varepsilon C_1C_2t},\\
& \pi=\underline{D_1}e^{\varepsilon x-(1-\varepsilon C_1C_2)t},\qquad \omega=\underline{D_2}e^{-\varepsilon x+(1-\varepsilon C_1C_2)t},\qquad U=C_2e^{\varepsilon x-t},\\ & V=C_1e^{t-\varepsilon x},
\end{split}
\label{solution8}
\end{equation}
where $\underline{D_1}\underline{D_2}=-C_1C_2$. Here, the fields $\eta$ and $\psi$ vary exponentially in time, while the fields $\pi$, $\omega$, $U$ and $V$ take the form of plane waves.

For subalgebra ${\mathcal L}_{13}$, we obtain the solution 
\begin{equation}
\begin{split}
& R=\varepsilon\mu,\qquad S=-\varepsilon\mu,\qquad \eta=\underline{\eta_0},\qquad \psi=-2\mu\underline{D_3}e^{\frac{1}{2}(\varepsilon t-\mu x)}+\underline{D_1},\\
& \pi=\underline{D_2}e^{\frac{1}{2}(\mu x+\varepsilon t)}+\frac{1}{2}C_1\underline{D_3}e^{\varepsilon t},\qquad \omega=0,\qquad U=C_1e^{\frac{1}{2}(\mu x+\varepsilon t)},\qquad V=0,
\end{split}
\label{solution4}
\end{equation}
which behaves similarly to the solution given in (\ref{solution8}), and represents plane waves.

In the case of subalgebra ${\mathcal L}_{15}$, we obtain the following two solutions:
\begin{equation}
\begin{split}
& R=\varepsilon,\qquad S=\varepsilon,\qquad \eta=\underline{D_1}f_1(x-\varepsilon t),\qquad \psi=\underline{D_1}f_2(x-\varepsilon t),\\
& \pi=\underline{D_3}f_3(x-\varepsilon t),\qquad \omega=\underline{D_3}f_4(x-\varepsilon t),\qquad U=\underline{D_1}\underline{D_3}g_1(x-\varepsilon t),\\ & V=\underline{D_1}\underline{D_3}g_2(x-\varepsilon t),
\end{split}
\label{solution7a}
\end{equation}
where $f_1$, $f_2$, $f_3$, $f_4$, $g_1$ and $g_2$ are arbitrary bosonic functions of $x-\varepsilon t$, and
\begin{equation}
\begin{split}
& R=\varepsilon,\qquad S=-\varepsilon,\qquad \eta=\underline{\eta_0},\qquad \psi=\psi(x-\varepsilon t),\\
& \pi=\underline{\pi_0},\qquad \omega=\omega(x-\varepsilon t),\qquad U=U_0,\qquad V=V(x-\varepsilon t),
\end{split}
\label{solution7b}
\end{equation}
Solutions (\ref{solution7a}) and (\ref{solution7b}) consist of travelling waves of arbitrary shape for both the bosonic and fermionic component fields. Bumps, kinks and doubly periodic solutions can be given a physical interpretation.

A solution invariant under subalgebra ${\mathcal L}_{68}$ can be written as
\begin{equation}
\begin{split}
& R=\varepsilon,\qquad S=-\varepsilon,\qquad \eta=-\frac{1}{4}\mu\underline{\alpha}g(x-\varepsilon t)+\frac{1}{2}\varepsilon\underline{\alpha}(x+\varepsilon t)+\underline{D_2},\\
& \psi=\underline{\alpha}f(x-\varepsilon t),\qquad \pi=-\frac{1}{4}\varepsilon\mu\underline{\alpha}f(x-\varepsilon t)+\underline{D_1},\qquad \omega=\underline{\alpha}g(x-\varepsilon t),\\
& U=\frac{1}{2}\varepsilon\mu(x+\varepsilon t),\qquad V=V_0,
\end{split}
\label{solution1}
\end{equation}
where $f$ and $g$ are arbitrary bosonic functions of $x-\varepsilon t$. This represents a combination of two travelling waves of arbitrary shape, including a superposition for the fermionic field $\eta$.

Finally, for subalgebra ${\mathcal L}_{149}$, we obtain the solution
\begin{equation}
\begin{split}
& R=\varepsilon,\qquad S=\varepsilon,\qquad \eta=\underline{\alpha}g(x-\varepsilon t),\qquad \psi=\varepsilon\underline{\alpha}\int{\mbox{d}\xi\over f_{\xi}}\bigg{|}_{\xi=x-\varepsilon t}+\underline{D_1},\\
& \pi={\mathcal P}(x-\varepsilon t)+\underline{\alpha}t,\qquad \omega=\underline{\omega_0},\qquad U=f(x-\varepsilon t),\qquad V=V_0,
\end{split}
\label{solution6}
\end{equation}
where $f$ and $g$ are arbitrary bosonic functions of $x-\varepsilon t$, and ${\mathcal P}$ is an arbitrary fermionic function of $x-\varepsilon t$. For certain specific functions $f$ (for instance, $f(\xi)=\xi^2$), the fermionic function $\psi$ admits the gradient catastrophe.

\section{Conclusion}

In this paper we have shown how to formulate an $N=2$ supersymmetric extension of a hydrodynamic-type system in Riemann invariants in terms of bosonic superfields. A Lie superalgebra of infinitesimal  symmetry generators of this extended system was found, consisting of translations and scaling transformations in both the bosonic and fermionic variables. In contrast with the $N=1$ case, more scaling transformations are present, while no boost-type generator was found. A systematic classification in terms of conjugacy classes was performed for the one-dimensional subalgebras, resulting in a list of 401 nonequivalent classes of subalgebras. It is interesting and significant to note that the classification is much more extensive for the $N=2$ supersymmetric extension than for its $N=1$ counterpart \cite{GrundHaritRiemann}. Consequently, a complete symmetry reduction analysis of our supersymmetric hydrodynamic system would lead to very large classes of invariant solutions. We illustrate this directly through several examples of new explicit solutions in closed form involving both bosonic and fermonic fields. These solutions include travelling waves, bumps, kinks, double-periodic solutions and solutions involving polynomials, exponentials and radicals. We have also demonstrated an example where the gradient catastrophe occurs for the supersymmetric version. In the classical case, bumps can be interpreted physically as nucleation centres, kinks as domain walls and nonsingular periodic solutions as elementary excitations \cite{Bishop,Tuszynski}. The question arises as to whether a similar physical interpretation can be made for the $N=2$ supersymmetric model.

The Lie symmetry algebra of the classical hydrodynamic system (\ref{b16}) shares in common with its classical Schr\"{o}dinger counterpart time and space translations together with a Galilean-type boost and a dilation in time and space \cite{GrundHaritRiemann,Unterberger}. The classical hydrodynamic algebra also contains a second dilation involving the fields and an inverse boost, while the classical Schr\"{o}dinger algebra contains a phase shift and a special conformal-type transformation. The phase shift is not present for the classical system (\ref{b16}) since the fields $R$ and $S$ appear explicitly in the equations.

The (12-dimensional) Lie superalgebra of the extended $N=2$ hydrodynamic system (\ref{e8}) and the (9-dimensional) superalgebra of the $N=2$ super-Schr\"{o}dinger model both contain translations in space and time plus an additional bosonic translation, a dilation and four fermionic translations (see Proposition 3.1 in \cite{Unterberger}). Also included in the supersymmetric hydrodynamic system's superalgebra is yet another bosonic translation and three additional dilations involving bosonic and fermionic variables, while the super-Schr\"{o}dinger superalgebra contains a supplementary boost-type transformation. The Lagrangian is known for the $N=2$ super-Schr\"{o}dinger model \cite{Unterberger} and one may inquire as to the nature of the Lagrangian formalism for the $N=2$ supersymmetric hydrodynamic model. Such an analysis may allow us to interpret the connection between the two systems and acquire a better understanding of the physical relevance of the supersymmetric hydrodynamic model in Riemann invariants.

\subsection*{Acknowledgements}

This project was completed during A.M.G.'s visit to the \'{E}cole Normale Sup\'{e}rieure de Cachan (Centre de Math\'{e}matiques et leurs Applications) and he would like to thank the CMLA for their kind invitation. This work was supported by research grants from NSERC of Canada.

\appendix

\section*{Appendix A: Subalgebras of the Lie superalgebra of the extended hydrodynamics system}

In the following discussion, we describe the classification of the Lie superalgebra described in Section 3. The symbols $\varepsilon$, $\mu$, and $\nu$ represent either $1$ or $-1$, the parameters $a$, $b$ and $c$ are nonzero real (bosonic) constants, while the quantities $\underline{\alpha}$, $\underline{\beta}$, $\underline{\gamma}$ and $\underline{\delta}$ are fermionic constants. The superalgebra ${\mathcal L}$ can be written in the composite semidirect sum given by (\ref{f2}).
In order to construct the list of representative one-dimensional subalgebras of ${\mathcal L}$, we begin by considering the subalgebras of the algebra ${\mathcal M}^{(0)}=\{M_1,M_2,M_3,M_4\}$. This algebra was classified by J. Patera and P. Winternitz as part of their general classification of real three- and four-dimensional Lie algebras \cite{Patera}. The list of representative subalgebras is
\begin{displaymath}
\begin{split}
{\mathcal L}_{1}&=\{M_4\},\quad {\mathcal L}_{2}=\{M_3+aM_4\},\quad {\mathcal L}_{3}=\{M_2+aM_3+bM_4\},\\ {\mathcal L}_{4}&=\{M_1+aM_2+bM_3+cM_4\}.
\end{split}
\end{displaymath}
We then proceed to determine the subalgebra classification of each successive composite semidirect sum in (\ref{f2}) using the procedure for semidirect sums of Lie algebras described in \cite{Winternitz}. This allows us to determine the subalgebra classification for each step and ultimately for the entire subalgebra ${\mathcal L}$, which to our knowledge has not been classified before.

\noindent For the algebra ${\mathcal M}^{(1)}={\mathcal M}^{(0)}\sdir\{P_0\}$, the splitting one-dimensional subalgebras consist of all the subalgebras determined above for ${\mathcal M}^{(0)}$ (i.e. ${\mathcal L}_{1}$, ${\mathcal L}_{2}$, ${\mathcal L}_{3}$ and ${\mathcal L}_{4}$) together with the subalgebra
\begin{displaymath}
{\mathcal L}_{5}=\{P_0\},
\end{displaymath}
and the nonsplitting one-dimensional subalgebras are
\begin{displaymath}
\begin{split}
{\mathcal L}_{6}&=\{M_4+\varepsilon P_0\},\quad {\mathcal L}_{7}=\{M_3+aM_4+\varepsilon P_0\},\quad {\mathcal L}_{8}=\{M_1+aM_3+bM_4+\varepsilon P_0\}.
\end{split}
\end{displaymath}

\noindent For the algebra ${\mathcal M}^{(2)}={\mathcal M}^{(1)}\sdir\{P_1\}$, the additional one-dimensional subalgebras are
\begin{displaymath}
\begin{split}
{\mathcal L}_{9}&=\{P_1\},\qquad {\mathcal L}_{10}=\{M_4+\varepsilon P_1\},\qquad {\mathcal L}_{11}=\{M_3+aM_4+\varepsilon P_1\},\\
{\mathcal L}_{12}&=\{M_2+aM_3+bM_4+\varepsilon P_1\},\quad {\mathcal L}_{13}=\{M_4+\varepsilon P_0+\mu P_1\},\\ {\mathcal L}_{14}&=\{M_3+aM_4+\varepsilon P_0+\mu P_1\},\quad {\mathcal L}_{15}=\{P_0+\varepsilon P_1\}.
\end{split}
\end{displaymath}

\noindent For the algebra ${\mathcal M}^{(3)}={\mathcal M}^{(2)}\sdir\{T_1\}$, the additional one-dimensional subalgebras are
\begin{displaymath}
\begin{split}
{\mathcal L}_{16}&=\{T_1\},\quad {\mathcal L}_{17}=\{M_3-M_4+\varepsilon T_1\},\quad
{\mathcal L}_{18}=\{M_2+aM_3+bM_4+\varepsilon T_1\}{\mbox{ (where }1+a+b=0)},\\ 
{\mathcal L}_{19}&=\{M_1+aM_2+bM_3+cM_4+\varepsilon T_1\}{\mbox{ (where }-1+a+b+c=0)},\quad {\mathcal L}_{20}=\{P_0+\varepsilon T_1\},\\ 
{\mathcal L}_{21}&=\{M_3-M_4+\varepsilon P_0+\mu T_1\},\quad {\mathcal L}_{22}=\{M_1+aM_3+bM_4+\varepsilon P_0+\mu T_1\}{\mbox{ (where }-1+a+b=0)},\\ {\mathcal L}_{23}&=\{P_1+\varepsilon T_1\},\quad {\mathcal L}_{24}=\{M_3-M_4+\varepsilon P_1+\mu T_1\},\\ 
{\mathcal L}_{25}&=\{M_2+aM_3+bM_4+\varepsilon P_1+\mu T_1\}{\mbox{ (where }1+a+b=0)},\\
{\mathcal L}_{26}&=\{M_3-M_4+\varepsilon P_0+\mu P_1+\nu T_1\},\quad
{\mathcal L}_{27}=\{P_0+\varepsilon P_1+\mu T_1\}.
\end{split}
\end{displaymath}

\noindent For the algebra ${\mathcal M}^{(4)}={\mathcal M}^{(3)}\sdir\{T_2\}$, the additional subalgebras are
\begin{displaymath}
\begin{split}
{\mathcal L}_{28}&=\{T_2\},\quad {\mathcal L}_{29}=\{M_3-M_4+\varepsilon T_2\},\quad
{\mathcal L}_{30}=\{M_2+aM_3+bM_4+\varepsilon T_2\}{\mbox{ (where }2+a+b=0)},\\ 
{\mathcal L}_{31}&=\{M_1+aM_2+bM_3+cM_4+\varepsilon T_2\}{\mbox{ (where }-4+2a+b+c=0)},\quad {\mathcal L}_{32}=\{P_0+\varepsilon T_2\},\\
{\mathcal L}_{33}&=\{M_3-M_4+\varepsilon P_0+\mu T_2\},\quad {\mathcal L}_{34}=\{M_1+aM_3+bM_4+\varepsilon P_0+\mu T_2\}{\mbox{ (where }4-a-b=0)},\\ {\mathcal L}_{35}&=\{P_1+\varepsilon T_2\},\quad {\mathcal L}_{36}=\{M_3-M_4+\varepsilon P_1+\mu T_2\},\\
\end{split}
\end{displaymath}
\begin{displaymath}
\begin{split}
{\mathcal L}_{37}&=\{M_2+aM_3+bM_4+\varepsilon P_1+\mu T_2\}{\mbox{ (where }2+a+b=0)},\quad {\mathcal L}_{38}=\{M_3-M_4+\varepsilon P_0+\mu P_1+\nu T_2\},\\ 
{\mathcal L}_{39}&=\{P_0+\varepsilon P_1+\mu T_2\},\quad {\mathcal L}_{40}=\{T_1+\varepsilon T_2\},\quad {\mathcal L}_{41}=\{M_3-M_4+\varepsilon T_1+\mu T_2\},\\ 
{\mathcal L}_{42}&=\{M_1+3M_2+bM_3+cM_4+\varepsilon T_1+\mu T_2\}{\mbox{ (where }2+b+c=0)},\quad {\mathcal L}_{43}=\{P_0+\varepsilon T_1+\mu T_2\},\\ {\mathcal L}_{44}&=\{M_3-M_4+\varepsilon P_0+\mu T_1+\nu T_2\},\quad {\mathcal L}_{45}=\{P_1+\varepsilon T_1+\mu T_2\},\\ 
{\mathcal L}_{46}&=\{M_3-M_4+\varepsilon P_1+\mu T_1+\nu T_2\},\quad {\mathcal L}_{47}=\{M_3-M_4+\varepsilon P_0+\mu P_1+\nu T_1+kT_2\}{\mbox{ (where }k\neq 0)},\\ 
{\mathcal L}_{48}&=\{P_0+\varepsilon P_1+\mu T_1+kT_2\}{\mbox{ (where }k\neq 0)}.
\end{split}
\end{displaymath}

\noindent For the superalgebra ${\mathcal M}^{(5)}={\mathcal M}^{(4)}\sdir\{Z_1\}$, the additional subalgebras are
\begin{displaymath}
\begin{split}
{\mathcal L}_{49}&=\{Z_1\},\quad {\mathcal L}_{50}=\{M_4+\underline{\alpha}Z_1\},\quad
{\mathcal L}_{51}=\{M_2+aM_4+\underline{\alpha}Z_1\},\\
{\mathcal L}_{52}&=\{M_1+aM_2+bM_4+\underline{\alpha}Z_1\},\quad {\mathcal L}_{53}=\{P_0+\underline{\alpha}Z_1\},\quad
{\mathcal L}_{54}=\{M_4+\varepsilon P_0+\underline{\alpha}Z_1\},\\ 
{\mathcal L}_{55}&=\{M_1+aM_4+\varepsilon P_0+\underline{\alpha}Z_1\},\quad {\mathcal L}_{56}=\{P_1+\underline{\alpha}Z_1\},\quad
{\mathcal L}_{57}=\{M_4+\varepsilon P_1+\underline{\alpha}Z_1\},\\ 
{\mathcal L}_{58}&=\{M_2+aM_4+\varepsilon P_1+\underline{\alpha}Z_1\},\quad {\mathcal L}_{59}=\{M_4+\varepsilon P_0+\mu P_1+\underline{\alpha}Z_1\},\quad
{\mathcal L}_{60}=\{P_0+\varepsilon P_1+\underline{\alpha}Z_1\},\\ 
{\mathcal L}_{61}&=\{T_1+\underline{\alpha}Z_1\},\quad {\mathcal L}_{62}=\{M_2-M_4+\varepsilon T_1+\underline{\alpha}Z_1\},\\
{\mathcal L}_{63}&=\{M_1+aM_2+bM_4+\varepsilon T_1+\underline{\alpha}Z_1\}{\mbox{ (where }-1+a+b=0)},\quad {\mathcal L}_{64}=\{P_0+\varepsilon T_1+\underline{\alpha}Z_1\},\\ 
{\mathcal L}_{65}&=\{M_1+M_4+\varepsilon P_0+\mu T_1+\underline{\alpha}Z_1\},\quad {\mathcal L}_{66}=\{P_1+\varepsilon T_1+\underline{\alpha}Z_1\},\\
{\mathcal L}_{67}&=\{M_2-M_4+\varepsilon P_1+\mu T_1+\underline{\alpha}Z_1\},\quad {\mathcal L}_{68}=\{P_0+\varepsilon P_1+\mu T_1+\underline{\alpha}Z_1\},\quad {\mathcal L}_{69}=\{T_2+\underline{\alpha}Z_1\},\\ 
{\mathcal L}_{70}&=\{M_2-2M_4+\varepsilon T_2+\underline{\alpha}Z_1\},\quad {\mathcal L}_{71}=\{M_1+aM_2+cM_4+\varepsilon T_2+\underline{\alpha}Z_1\}{\mbox{ (where }-4+2a+c=0)},\\ 
{\mathcal L}_{72}&=\{P_0+\varepsilon T_2+\underline{\alpha}Z_1\},\quad {\mathcal L}_{73}=\{M_1+4M_4+\varepsilon P_0+\mu T_2+\underline{\alpha}Z_1\},\quad {\mathcal L}_{74}=\{P_1+\varepsilon T_2+\underline{\alpha}Z_1\},\\ 
{\mathcal L}_{75}&=\{M_2-2M_4+\varepsilon P_1+\mu T_2+\underline{\alpha}Z_1\},\quad {\mathcal L}_{76}=\{P_0+\varepsilon P_1+\mu T_2+\underline{\alpha}Z_1\},\\ 
{\mathcal L}_{77}&=\{T_1+\varepsilon T_2+\underline{\alpha}Z_1\},\quad {\mathcal L}_{78}=\{M_1+3M_2-2M_4+\varepsilon T_1+\mu T_2+\underline{\alpha}Z_1\},\\
{\mathcal L}_{79}&=\{P_0+\varepsilon T_1+\mu T_2+\underline{\alpha}Z_1\},\quad {\mathcal L}_{80}=\{P_1+\varepsilon T_1+\mu T_2+\underline{\alpha}Z_1\},\\ {\mathcal L}_{81}&=\{P_0+\varepsilon P_1+\mu T_1+kT_2+\underline{\alpha}Z_1\}{\mbox{ (where }k\neq 0)}.
\end{split}
\end{displaymath}

\noindent For the superalgebra ${\mathcal M}^{(6)}={\mathcal M}^{(5)}\sdir\{Z_2\}$, the additional subalgebras are
\begin{displaymath}
\begin{split}
{\mathcal L}_{82}&=\{Z_2\},\quad {\mathcal L}_{83}=\{M_4+\underline{\alpha}Z_2\},\quad
{\mathcal L}_{84}=\{M_2-M_3+bM_4+\underline{\alpha}Z_2\},\\ 
{\mathcal L}_{85}&=\{M_1+aM_2+bM_3+cM_4+\underline{\alpha}Z_2\}{\mbox{ (where }2-a-b=0)},\quad {\mathcal L}_{86}=\{P_0+\underline{\alpha}Z_2\},\\ 
{\mathcal L}_{87}&=\{M_4+\varepsilon P_0+\underline{\alpha}Z_2\},\quad {\mathcal L}_{88}=\{M_1+2M_3+bM_4+\varepsilon P_0+\underline{\alpha}Z_2\},\quad
{\mathcal L}_{89}=\{P_1+\underline{\alpha}Z_2\},\\ 
{\mathcal L}_{90}&=\{M_4+\varepsilon P_1+\underline{\alpha}Z_2\},\quad {\mathcal L}_{91}=\{M_2-M_3+bM_4+\varepsilon P_1+\underline{\alpha}Z_2\},\\
{\mathcal L}_{92}&=\{M_4+\varepsilon P_0+\mu P_1+\underline{\alpha}Z_2\},\quad {\mathcal L}_{93}=\{P_0+\varepsilon P_1+\underline{\alpha}Z_2\},\quad {\mathcal L}_{94}=\{T_1+\underline{\alpha}Z_2\},\\
{\mathcal L}_{95}&=\{M_2-M_3+\varepsilon T_1+\underline{\alpha}Z_2\},\quad {\mathcal L}_{96}=\{M_1+aM_2+bM_3-M_4+\varepsilon T_1+\underline{\alpha}Z_2\}{\mbox{ (where }2-a-b=0)},\\ 
\end{split}
\end{displaymath}
\begin{displaymath}
\begin{split}
{\mathcal L}_{97}&=\{P_0+\varepsilon T_1+\underline{\alpha}Z_2\},\quad {\mathcal L}_{98}=\{M_1+2M_3-M_4+\varepsilon P_0+\mu T_1+\underline{\alpha}Z_2\},\\ {\mathcal L}_{99}&=\{P_1+\varepsilon T_1+\underline{\alpha}Z_2\},\quad {\mathcal L}_{100}=\{M_2-M_3+\varepsilon P_1+\mu T_1+\underline{\alpha}Z_2\},\\ {\mathcal L}_{101}&=\{P_0+\varepsilon P_1+\mu T_1+\underline{\alpha}Z_2\},\quad {\mathcal L}_{102}=\{T_2+\underline{\alpha}Z_2\},\quad {\mathcal L}_{103}=\{M_2-M_3-M_4+\varepsilon T_2+\underline{\alpha}Z_2\},\\ 
{\mathcal L}_{104}&=\{M_1+aM_2+bM_3+bM_4+\varepsilon T_2+\underline{\alpha}Z_2\}{\mbox{ (where }2-a-b=0)},\quad {\mathcal L}_{105}=\{P_0+\varepsilon T_2+\underline{\alpha}Z_2\},\\ 
{\mathcal L}_{106}&=\{M_1+2M_3+2M_4+\varepsilon P_0+\mu T_2+\underline{\alpha}Z_2\},\quad {\mathcal L}_{107}=\{P_1+\varepsilon T_2+\underline{\alpha}Z_2\},\\ 
{\mathcal L}_{108}&=\{M_2-M_3-M_4+\varepsilon P_1+\mu T_2+\underline{\alpha}Z_2\},\quad {\mathcal L}_{109}=\{P_0+\varepsilon P_1+\mu T_2+\underline{\alpha}Z_2\},\\ 
{\mathcal L}_{110}&=\{T_1+\varepsilon T_2+\underline{\alpha}Z_2\},\quad {\mathcal L}_{111}=\{M_1+3M_2-M_3-M_4+\varepsilon T_1+\mu T_2+\underline{\alpha}Z_2\},\\ 
{\mathcal L}_{112}&=\{P_0+\varepsilon T_1+\mu T_2+\underline{\alpha}Z_2\},\quad {\mathcal L}_{113}=\{P_1+\varepsilon T_1+\mu T_2+\underline{\alpha}Z_2\},\\ {\mathcal L}_{114}&=\{P_0+\varepsilon P_1+\mu T_1+kT_2+\underline{\alpha}Z_2\}{\mbox{ (where }k\neq 0)},\quad {\mathcal L}_{115}=\{Z_1+kZ_2\}{\mbox{ (where }k\neq 0)},\\ 
{\mathcal L}_{116}&=\{M_4+\underline{\alpha}Z_1+\underline{\beta}Z_2\},\quad {\mathcal L}_{117}=\{M_1+2M_2+bM_4+\underline{\alpha}Z_1+\underline{\beta}Z_2\},\\ 
{\mathcal L}_{118}&=\{P_0+\underline{\alpha}Z_1+\underline{\beta}Z_2\},\quad {\mathcal L}_{119}=\{M_4+\varepsilon P_0+\underline{\alpha}Z_1+\underline{\beta}Z_2\},\quad {\mathcal L}_{120}=\{P_1+\underline{\alpha}Z_1+\underline{\beta}Z_2\},\\
{\mathcal L}_{121}&=\{M_4+\varepsilon P_1+\underline{\alpha}Z_1+\underline{\beta}Z_2\},\quad {\mathcal L}_{122}=\{M_4+\varepsilon P_0+\mu P_1+\underline{\alpha}Z_1+\underline{\beta}Z_2\},\\ 
{\mathcal L}_{123}&=\{P_0+\varepsilon P_1+\underline{\alpha}Z_1+\underline{\beta}Z_2\},\quad {\mathcal L}_{124}=\{T_1+\underline{\alpha}Z_1+\underline{\beta}Z_2\},\\ 
{\mathcal L}_{125}&=\{M_1+2M_2-M_4+\varepsilon T_1+\underline{\alpha}Z_1+\underline{\beta}Z_2\},\quad {\mathcal L}_{126}=\{P_0+\varepsilon T_1+\underline{\alpha}Z_1+\underline{\beta}Z_2\},\\
{\mathcal L}_{127}&=\{P_1+\varepsilon T_1+\underline{\alpha}Z_1+\underline{\beta}Z_2\},\quad {\mathcal L}_{128}=\{P_0+\varepsilon P_1+\mu T_1+\underline{\alpha}Z_1+\underline{\beta}Z_2\},\\ 
{\mathcal L}_{129}&=\{T_2+\underline{\alpha}Z_1+\underline{\beta}Z_2\},\quad {\mathcal L}_{130}=\{M_1+2M_2+\varepsilon T_2+\underline{\alpha}Z_1+\underline{\beta}Z_2\},\\ 
{\mathcal L}_{131}&=\{P_0+\varepsilon T_2+\underline{\alpha}Z_1+\underline{\beta}Z_2\},\quad {\mathcal L}_{132}=\{P_1+\varepsilon T_2+\underline{\alpha}Z_1+\underline{\beta}Z_2\},\\
{\mathcal L}_{133}&=\{P_0+\varepsilon P_1+\mu T_2+\underline{\alpha}Z_1+\underline{\beta}Z_2\},\quad {\mathcal L}_{134}=\{T_1+\varepsilon T_2+\underline{\alpha}Z_1+\underline{\beta}Z_2\},\\ 
{\mathcal L}_{135}&=\{P_0+\varepsilon T_1+\mu T_2+\underline{\alpha}Z_1+\underline{\beta}Z_2\},\quad {\mathcal L}_{136}=\{P_1+\varepsilon T_1+\mu T_2+\underline{\alpha}Z_1+\underline{\beta}Z_2\},\\ 
{\mathcal L}_{137}&=\{P_0+\varepsilon P_1+\mu T_1+kT_2+\underline{\alpha}Z_1+\underline{\beta}Z_2\}{\mbox{ (where }k\neq 0)}.
\end{split}
\end{displaymath}

\noindent For the superalgebra ${\mathcal M}^{(7)}={\mathcal M}^{(6)}\sdir\{Z_3\}$, the additional subalgebras are
\begin{displaymath}
\begin{split}
{\mathcal L}_{138}&=\{Z_3\},\quad {\mathcal L}_{139}=\{M_3+\underline{\alpha}Z_3\},\quad
{\mathcal L}_{140}=\{M_2+aM_3+\underline{\alpha}Z_3\},\\ 
{\mathcal L}_{141}&=\{M_1+aM_2+bM_3+\underline{\alpha}Z_3\},\quad {\mathcal L}_{142}=\{P_0+\underline{\alpha}Z_3\},\quad
{\mathcal L}_{143}=\{M_3+\varepsilon P_0+\underline{\alpha}Z_3\},\\ 
{\mathcal L}_{144}&=\{M_1+aM_3+\varepsilon P_0+\underline{\alpha}Z_3\},\quad {\mathcal L}_{145}=\{P_1+\underline{\alpha}Z_3\},\quad
{\mathcal L}_{146}=\{M_3+\varepsilon P_1+\underline{\alpha}Z_3\},\\ 
{\mathcal L}_{147}&=\{M_2+aM_3+\varepsilon P_1+\underline{\alpha}Z_3\},\quad
{\mathcal L}_{148}=\{M_3+\varepsilon P_0+\mu P_1+\underline{\alpha}Z_3\},\\ 
{\mathcal L}_{149}&=\{P_0+\varepsilon P_1+\underline{\alpha}Z_3\},\quad {\mathcal L}_{150}=\{T_1+\underline{\alpha}Z_3\},\quad
{\mathcal L}_{151}=\{M_2-M_3+\varepsilon T_1+\underline{\alpha}Z_3\},\\ 
{\mathcal L}_{152}&=\{M_1+aM_2+bM_3+\varepsilon T_1+\underline{\alpha}Z_3\}{\mbox{ (where }-1+a+b=0)},\quad
{\mathcal L}_{153}=\{P_0+\varepsilon T_1+\underline{\alpha}Z_3\},\\
{\mathcal L}_{154}&=\{M_1+M_3+\varepsilon P_0+\mu T_1+\underline{\alpha}Z_3\},\quad {\mathcal L}_{155}=\{P_1+\varepsilon T_1+\underline{\alpha}Z_3\},\\ 
{\mathcal L}_{156}&=\{M_2-M_3+\varepsilon P_1+\mu T_1+\underline{\alpha}Z_3\},\quad {\mathcal L}_{157}=\{P_0+\varepsilon P_1+\mu T_1+\underline{\alpha}Z_3\},\quad {\mathcal L}_{158}=\{T_2+\underline{\alpha}Z_3\},\\
\end{split}
\end{displaymath}
\begin{displaymath}
\begin{split}
{\mathcal L}_{159}&=\{M_2-2M_3+\varepsilon T_2+\underline{\alpha}Z_3\},\quad {\mathcal L}_{160}=\{M_1+aM_2+bM_3+\varepsilon T_2+\underline{\alpha}Z_3\}{\mbox{ (where }-4+2a+b=0)},\\ 
{\mathcal L}_{161}&=\{P_0+\varepsilon T_2+\underline{\alpha}Z_3\},\quad {\mathcal L}_{162}=\{M_1+4M_3+\varepsilon P_0+\mu T_2+\underline{\alpha}Z_3\},\quad {\mathcal L}_{163}=\{P_1+\varepsilon T_2+\underline{\alpha}Z_3\},\\ 
{\mathcal L}_{164}&=\{M_2-2M_3+\varepsilon P_1+\mu T_2+\underline{\alpha}Z_3\},\quad {\mathcal L}_{165}=\{P_0+\varepsilon P_1+\mu T_2+\underline{\alpha}Z_3\},\\ 
{\mathcal L}_{166}&=\{T_1+\varepsilon T_2+\underline{\alpha}Z_3\},\quad {\mathcal L}_{167}=\{M_1+3M_2-2M_3+\varepsilon T_1+\mu T_2+\underline{\alpha}Z_3\},\\ 
{\mathcal L}_{168}&=\{P_0+\varepsilon T_1+\mu T_2+\underline{\alpha}Z_3\},\quad {\mathcal L}_{169}=\{P_1+\varepsilon T_1+\mu T_2+\underline{\alpha}Z_3\},\\
{\mathcal L}_{170}&=\{P_0+\varepsilon P_1+\mu T_1+kT_2+\underline{\alpha}Z_3\}{\mbox{ (where }k\neq 0)},\quad {\mathcal L}_{171}=\{Z_1+kZ_3\}{\mbox{ (where }k\neq 0)},\\ 
{\mathcal L}_{172}&=\{M_2+\underline{\alpha}Z_1+\underline{\beta}Z_3\},\quad {\mathcal L}_{173}=\{M_1+aM_2+\underline{\alpha}Z_1+\underline{\beta}Z_3\},\quad {\mathcal L}_{174}=\{P_0+\underline{\alpha}Z_1+\underline{\beta}Z_3\},\\
{\mathcal L}_{175}&=\{M_1+\varepsilon P_0+\underline{\alpha}Z_1+\underline{\beta}Z_3\},\quad {\mathcal L}_{176}=\{P_1+\underline{\alpha}Z_1+\underline{\beta}Z_3\},\\ 
{\mathcal L}_{177}&=\{M_2+\varepsilon P_1+\underline{\alpha}Z_1+\underline{\beta}Z_3\},\quad {\mathcal L}_{178}=\{P_0+\varepsilon P_1+\underline{\alpha}Z_1+\underline{\beta}Z_3\},\\ 
{\mathcal L}_{179}&=\{T_1+\underline{\alpha}Z_1+\underline{\beta}Z_3\},\quad {\mathcal L}_{180}=\{M_1+M_2+\varepsilon T_1+\underline{\alpha}Z_1+\underline{\beta}Z_3\},\\ 
{\mathcal L}_{181}&=\{P_0+\varepsilon T_1+\underline{\alpha}Z_1+\underline{\beta}Z_3\},\quad {\mathcal L}_{182}=\{P_1+\varepsilon T_1+\underline{\alpha}Z_1+\underline{\beta}Z_3\},\\ 
{\mathcal L}_{183}&=\{P_0+\varepsilon P_1+\mu T_1+\underline{\alpha}Z_1+\underline{\beta}Z_3\},\quad {\mathcal L}_{184}=\{T_2+\underline{\alpha}Z_1+\underline{\beta}Z_3\},\\ 
{\mathcal L}_{185}&=\{M_1+2M_2+\varepsilon T_2+\underline{\alpha}Z_1+\underline{\beta}Z_3\},\quad {\mathcal L}_{186}=\{P_0+\varepsilon T_2+\underline{\alpha}Z_1+\underline{\beta}Z_3\},\\
{\mathcal L}_{187}&=\{P_1+\varepsilon T_2+\underline{\alpha}Z_1+\underline{\beta}Z_3\},\quad {\mathcal L}_{188}=\{P_0+\varepsilon P_1+\mu T_2+\underline{\alpha}Z_1+\underline{\beta}Z_3\},\\ 
{\mathcal L}_{189}&=\{T_1+\varepsilon T_2+\underline{\alpha}Z_1+\underline{\beta}Z_3\},\quad {\mathcal L}_{190}=\{P_0+\varepsilon T_1+\mu T_2+\underline{\alpha}Z_1+\underline{\beta}Z_3\},\\ 
{\mathcal L}_{191}&=\{P_1+\varepsilon T_1+\mu T_2+\underline{\alpha}Z_1+\underline{\beta}Z_3\},\quad {\mathcal L}_{192}=\{P_0+\varepsilon P_1+\mu T_1+kT_2+\underline{\alpha}Z_1+\underline{\beta}Z_3\}{\mbox{ (where }k\neq 0)},\\
{\mathcal L}_{193}&=\{Z_2+kZ_3\}{\mbox{ (where }k\neq 0)},\quad {\mathcal L}_{194}=\{M_2-M_3+\underline{\alpha}Z_2+\underline{\beta}Z_3\},\\ 
{\mathcal L}_{195}&=\{M_1+aM_2+bM_3+\underline{\alpha}Z_2+\underline{\beta}Z_3\}{\mbox{ (where }2-a-b=0)},\quad {\mathcal L}_{196}=\{P_0+\underline{\alpha}Z_2+\underline{\beta}Z_3\},\\ 
{\mathcal L}_{197}&=\{M_1+2M_3+\varepsilon P_0+\underline{\alpha}Z_2+\underline{\beta}Z_3\},\quad {\mathcal L}_{198}=\{P_1+\underline{\alpha}Z_2+\underline{\beta}Z_3\},\\
{\mathcal L}_{199}&=\{M_2-M_3+\varepsilon P_1+\underline{\alpha}Z_2+\underline{\beta}Z_3\},\quad {\mathcal L}_{200}=\{P_0+\varepsilon P_1+\underline{\alpha}Z_2+\underline{\beta}Z_3\},\\ 
{\mathcal L}_{201}&=\{T_1+\underline{\alpha}Z_2+\underline{\beta}Z_3\},\quad {\mathcal L}_{202}=\{M_2-M_3+\varepsilon T_1+\underline{\alpha}Z_2+\underline{\beta}Z_3\},\\ 
{\mathcal L}_{203}&=\{P_0+\varepsilon T_1+\underline{\alpha}Z_2+\underline{\beta}Z_3\},\quad {\mathcal L}_{204}=\{P_1+\varepsilon T_1+\underline{\alpha}Z_2+\underline{\beta}Z_3\},\\
{\mathcal L}_{205}&=\{M_2-M_3+\varepsilon P_1+\mu T_1+\underline{\alpha}Z_2+\underline{\beta}Z_3\},\quad {\mathcal L}_{206}=\{P_0+\varepsilon P_1+\mu T_1+\underline{\alpha}Z_2+\underline{\beta}Z_3\},\\ 
{\mathcal L}_{207}&=\{T_2+\underline{\alpha}Z_2+\underline{\beta}Z_3\},\quad {\mathcal L}_{208}=\{M_1+2M_2+\varepsilon T_2+\underline{\alpha}Z_2+\underline{\beta}Z_3\},\\ 
{\mathcal L}_{209}&=\{P_0+\varepsilon T_2+\underline{\alpha}Z_2+\underline{\beta}Z_3\},\quad {\mathcal L}_{210}=\{P_1+\varepsilon T_2+\underline{\alpha}Z_2+\underline{\beta}Z_3\},\\
{\mathcal L}_{211}&=\{P_0+\varepsilon P_1+\mu T_2+\underline{\alpha}Z_2+\underline{\beta}Z_3\},\quad {\mathcal L}_{212}=\{T_1+\varepsilon T_2+\underline{\alpha}Z_2+\underline{\beta}Z_3\},\\ 
{\mathcal L}_{213}&=\{P_0+\varepsilon T_1+\mu T_2+\underline{\alpha}Z_2+\underline{\beta}Z_3\},\quad {\mathcal L}_{214}=\{P_1+\varepsilon T_1+\mu T_2+\underline{\alpha}Z_2+\underline{\beta}Z_3\},\\ 
{\mathcal L}_{215}&=\{P_0+\varepsilon P_1+\mu T_1+kT_2+\underline{\alpha}Z_2+\underline{\beta}Z_3\}{\mbox{ (where }k\neq 0)},\quad {\mathcal L}_{216}=\{Z_1+kZ_2+lZ_3\}{\mbox{ (where }k,l\neq 0)},\\
{\mathcal L}_{217}&=\{M_1+2M_2+\underline{\alpha}Z_1+\underline{\beta}Z_2+\underline{\gamma}Z_3\},\quad {\mathcal L}_{218}=\{P_0+\underline{\alpha}Z_1+\underline{\beta}Z_2+\underline{\gamma}Z_3\},\\
\end{split}
\end{displaymath}
\begin{displaymath}
\begin{split}
{\mathcal L}_{219}&=\{P_1+\underline{\alpha}Z_1+\underline{\beta}Z_2+\underline{\gamma}Z_3\},\quad {\mathcal L}_{220}=\{P_0+\varepsilon P_1+\underline{\alpha}Z_1+\underline{\beta}Z_2+\underline{\gamma}Z_3\},\\
{\mathcal L}_{221}&=\{T_1+\underline{\alpha}Z_1+\underline{\beta}Z_2+\underline{\gamma}Z_3\},\quad {\mathcal L}_{222}=\{P_0+\varepsilon T_1+\underline{\alpha}Z_1+\underline{\beta}Z_2+\underline{\gamma}Z_3\},\\
{\mathcal L}_{223}&=\{P_1+\varepsilon T_1+\underline{\alpha}Z_1+\underline{\beta}Z_2+\underline{\gamma}Z_3\},\quad {\mathcal L}_{224}=\{P_0+\varepsilon P_1+\mu T_1+\underline{\alpha}Z_1+\underline{\beta}Z_2+\underline{\gamma}Z_3\},\\ 
{\mathcal L}_{225}&=\{T_2+\underline{\alpha}Z_1+\underline{\beta}Z_2+\underline{\gamma}Z_3\},\quad {\mathcal L}_{226}=\{M_1+2M_2+\varepsilon T_2+\underline{\alpha}Z_1+\underline{\beta}Z_2+\underline{\gamma}Z_3\},\\ 
{\mathcal L}_{227}&=\{P_0+\varepsilon T_2+\underline{\alpha}Z_1+\underline{\beta}Z_2+\underline{\gamma}Z_3\},\quad {\mathcal L}_{228}=\{P_1+\varepsilon T_2+\underline{\alpha}Z_1+\underline{\beta}Z_2+\underline{\gamma}Z_3\},\\
{\mathcal L}_{229}&=\{P_0+\varepsilon P_1+\mu T_2+\underline{\alpha}Z_1+\underline{\beta}Z_2+\underline{\gamma}Z_3\},\quad {\mathcal L}_{230}=\{T_1+\varepsilon T_2+\underline{\alpha}Z_1+\underline{\beta}Z_2+\underline{\gamma}Z_3\},\\ 
{\mathcal L}_{231}&=\{P_0+\varepsilon T_1+\mu T_2+\underline{\alpha}Z_1+\underline{\beta}Z_2+\underline{\gamma}Z_3\},\quad {\mathcal L}_{232}=\{P_1+\varepsilon T_1+\mu T_2+\underline{\alpha}Z_1+\underline{\beta}Z_2+\underline{\gamma}Z_3\},\\ 
{\mathcal L}_{233}&=\{P_0+\varepsilon P_1+\mu T_1+kT_2+\underline{\alpha}Z_1+\underline{\beta}Z_2+\underline{\gamma}Z_3\}{\mbox{ (where }k\neq 0)}.
\end{split}
\end{displaymath}

\noindent For the full superalgebra ${\mathcal L}={\mathcal M}^{(8)}={\mathcal M}^{(7)}\sdir\{Z_4\}$, the additional subalgebras are
\begin{displaymath}
\begin{split}
{\mathcal L}_{234}&=\{Z_4\},\quad {\mathcal L}_{235}=\{M_3+\underline{\alpha}Z_4\},\quad {\mathcal L}_{236}=\{M_2+aM_3-2M_4+\underline{\alpha}Z_4\},\\ {\mathcal L}_{237}&=\{M_1+aM_2+bM_3+cM_4+\underline{\alpha}Z_4\}{\mbox{ (where }3-2a-c=0)},\quad {\mathcal L}_{238}=\{P_0+\underline{\alpha}Z_4\},\\ {\mathcal L}_{239}&=\{M_3+\varepsilon P_0+\underline{\alpha}Z_4\},\quad {\mathcal L}_{240}=\{M_1+aM_3+3M_4+\varepsilon P_0+\underline{\alpha}Z_4\},\quad
{\mathcal L}_{241}=\{P_1+\underline{\alpha}Z_4\},\\ 
{\mathcal L}_{242}&=\{M_3+\varepsilon P_1+\underline{\alpha}Z_4\},\quad {\mathcal L}_{243}=\{M_2+aM_3-2M_4+\varepsilon P_1+\underline{\alpha}Z_4\},\\ {\mathcal L}_{244}&=\{M_3+\varepsilon P_0+\mu P_1+\underline{\alpha}Z_4\},\quad {\mathcal L}_{245}=\{P_0+\varepsilon P_1+\underline{\alpha}Z_4\},\quad {\mathcal L}_{246}=\{T_1+\underline{\alpha}Z_4\},\\
{\mathcal L}_{247}&=\{M_2+M_3-2M_4+\varepsilon T_1+\underline{\alpha}Z_4\},\\ 
{\mathcal L}_{248}&=\{M_1+aM_2+(a-2)M_3+(3-2a)M_4+\varepsilon T_1+\underline{\alpha}Z_4\},\quad {\mathcal L}_{249}=\{P_0+\varepsilon T_1+\underline{\alpha}Z_4\},\\ 
{\mathcal L}_{250}&=\{M_1-2M_3+3M_4+\varepsilon P_0+\mu T_1+\underline{\alpha}Z_4\},\quad {\mathcal L}_{251}=\{P_1+\varepsilon T_1+\underline{\alpha}Z_4\},\\ 
{\mathcal L}_{252}&=\{M_2+M_3-2M_4+\varepsilon P_1+\mu T_1+\underline{\alpha}Z_4\},\quad {\mathcal L}_{253}=\{P_0+\varepsilon P_1+\mu T_1+\underline{\alpha}Z_4\},\\ 
{\mathcal L}_{254}&=\{T_2+\underline{\alpha}Z_4\},\quad {\mathcal L}_{255}=\{M_2-2M_4+\varepsilon T_2+\underline{\alpha}Z_4\},\\ 
{\mathcal L}_{256}&=\{M_1+aM_2+M_3+(3-2a)M_4+\varepsilon T_2+\underline{\alpha}Z_4\},\quad {\mathcal L}_{257}=\{P_0+\varepsilon T_2+\underline{\alpha}Z_4\},\\ 
{\mathcal L}_{258}&=\{M_1+M_3+3M_4+\varepsilon P_0+\mu T_2+\underline{\alpha}Z_4\},\quad
{\mathcal L}_{259}=\{P_1+\varepsilon T_2+\underline{\alpha}Z_4\},\\ 
{\mathcal L}_{260}&=\{M_2-2M_4+\varepsilon P_1+\mu T_2+\underline{\alpha}Z_4\},\quad {\mathcal L}_{261}=\{P_0+\varepsilon P_1+\mu T_2+\underline{\alpha}Z_4\},\\ 
{\mathcal L}_{262}&=\{T_1+\varepsilon T_2+\underline{\alpha}Z_4\},\quad {\mathcal L}_{263}=\{M_1+3M_2+M_3-3M_4+\varepsilon T_1+\mu T_2+\underline{\alpha}Z_4\},\\ 
{\mathcal L}_{264}&=\{P_0+\varepsilon T_1+\mu T_2+\underline{\alpha}Z_4\},\quad
{\mathcal L}_{265}=\{P_1+\varepsilon T_1+\mu T_2+\underline{\alpha}Z_4\},\\ 
{\mathcal L}_{266}&=\{P_0+\varepsilon P_1+\mu T_1+kT_2+\underline{\alpha}Z_4\}{\mbox{ (where }k\neq 0)},\quad {\mathcal L}_{267}=\{Z_1+kZ_4\}{\mbox{ (where }k\neq 0)},\\
{\mathcal L}_{268}&=\{M_2-2M_4+\underline{\alpha}Z_1+\underline{\beta}Z_4\},\quad {\mathcal L}_{269}=\{M_1+aM_2+bM_4+\underline{\alpha}Z_1+\underline{\beta}Z_4\}{\mbox{ (where }3-2a-b=0)},\\ 
{\mathcal L}_{270}&=\{P_0+\underline{\alpha}Z_1+\underline{\beta}Z_4\},\quad
{\mathcal L}_{271}=\{M_1+3M_4+\varepsilon P_0+\underline{\alpha}Z_1+\underline{\beta}Z_4\},\\ 
{\mathcal L}_{272}&=\{P_1+\underline{\alpha}Z_1+\underline{\beta}Z_4\},\quad {\mathcal L}_{273}=\{M_2-2M_4+\varepsilon P_1+\underline{\alpha}Z_1+\underline{\beta}Z_4\},\\ 
{\mathcal L}_{274}&=\{P_0+\varepsilon P_1+\underline{\alpha}Z_1+\underline{\beta}Z_4\},\quad {\mathcal L}_{275}=\{T_1+\underline{\alpha}Z_1+\underline{\beta}Z_4\},\\
\end{split}
\end{displaymath}
\begin{displaymath}
\begin{split}
{\mathcal L}_{276}&=\{M_1+2M_2-M_4+\varepsilon T_1+\underline{\alpha}Z_1+\underline{\beta}Z_4\},\quad
{\mathcal L}_{277}=\{P_0+\varepsilon T_1+\underline{\alpha}Z_1+\underline{\beta}Z_4\},\\ 
{\mathcal L}_{278}&=\{P_1+\varepsilon T_1+\underline{\alpha}Z_1+\underline{\beta}Z_4\},\quad {\mathcal L}_{279}=\{P_0+\varepsilon P_1+\mu T_1+\underline{\alpha}Z_1+\underline{\beta}Z_4\},\\ 
{\mathcal L}_{280}&=\{T_2+\underline{\alpha}Z_1+\underline{\beta}Z_4\},\quad {\mathcal L}_{281}=\{M_2-2M_4+\varepsilon T_2+\underline{\alpha}Z_1+\underline{\beta}Z_4\},\\ 
{\mathcal L}_{282}&=\{P_0+\varepsilon T_2+\underline{\alpha}Z_1+\underline{\beta}Z_4\},\quad
{\mathcal L}_{283}=\{P_1+\varepsilon T_2+\underline{\alpha}Z_1+\underline{\beta}Z_4\},\\ 
{\mathcal L}_{284}&=\{M_2-2M_4+\varepsilon P_1+\mu T_2+\underline{\alpha}Z_1+\underline{\beta}Z_4\},\quad {\mathcal L}_{285}=\{P_0+\varepsilon P_1+\mu T_2+\underline{\alpha}Z_1+\underline{\beta}Z_4\},\\ 
{\mathcal L}_{286}&=\{T_1+\varepsilon T_2+\underline{\alpha}Z_1+\underline{\beta}Z_4\},\quad {\mathcal L}_{287}=\{P_0+\varepsilon T_1+\mu T_2+\underline{\alpha}Z_1+\underline{\beta}Z_4\},\\ 
{\mathcal L}_{288}&=\{P_1+\varepsilon T_1+\mu T_2+\underline{\alpha}Z_1+\underline{\beta}Z_4\},\\
{\mathcal L}_{289}&=\{P_0+\varepsilon P_1+\mu T_1+kT_2+\underline{\alpha}Z_1+\underline{\beta}Z_4\}{\mbox{ (where }k\neq 0)},\quad {\mathcal L}_{290}=\{Z_2+kZ_4\}{\mbox{ (where }k\neq 0)},\\ 
{\mathcal L}_{291}&=\{M_1+aM_2+(2-a)M_3+(3-2a)M_4+\underline{\alpha}Z_2+\underline{\beta}Z_4\},\quad {\mathcal L}_{292}=\{P_0+\underline{\alpha}Z_2+\underline{\beta}Z_4\},\\ 
{\mathcal L}_{293}&=\{M_1+2M_3+3M_4+\varepsilon P_0+\underline{\alpha}Z_2+\underline{\beta}Z_4\},\quad {\mathcal L}_{294}=\{P_1+\underline{\alpha}Z_2+\underline{\beta}Z_4\},\\
{\mathcal L}_{295}&=\{M_2-M_3-2M_4+\varepsilon P_1+\underline{\alpha}Z_2+\underline{\beta}Z_4\},\quad {\mathcal L}_{296}=\{P_0+\varepsilon P_1+\underline{\alpha}Z_2+\underline{\beta}Z_4\},\\ 
{\mathcal L}_{297}&=\{T_1+\underline{\alpha}Z_2+\underline{\beta}Z_4\},\quad {\mathcal L}_{298}=\{M_1+2M_2-M_4+\varepsilon T_1+\underline{\alpha}Z_2+\underline{\beta}Z_4\},\\ 
{\mathcal L}_{299}&=\{P_0+\varepsilon T_1+\underline{\alpha}Z_2+\underline{\beta}Z_4\},\quad {\mathcal L}_{300}=\{P_1+\varepsilon T_1+\underline{\alpha}Z_2+\underline{\beta}Z_4\},\\
{\mathcal L}_{301}&=\{P_0+\varepsilon P_1+\mu T_1+\underline{\alpha}Z_2+\underline{\beta}Z_4\},\quad {\mathcal L}_{302}=\{T_2+\underline{\alpha}Z_2+\underline{\beta}Z_4\},\\ 
{\mathcal L}_{303}&=\{M_1+M_2+M_3+M_4+\varepsilon T_2+\underline{\alpha}Z_2+\underline{\beta}Z_4\},\quad {\mathcal L}_{304}=\{P_0+\varepsilon T_2+\underline{\alpha}Z_2+\underline{\beta}Z_4\},\\ 
{\mathcal L}_{305}&=\{P_1+\varepsilon T_2+\underline{\alpha}Z_2+\underline{\beta}Z_4\},\quad {\mathcal L}_{306}=\{P_0+\varepsilon P_1+\mu T_2+\underline{\alpha}Z_2+\underline{\beta}Z_4\},\\
{\mathcal L}_{307}&=\{T_1+\varepsilon T_2+\underline{\alpha}Z_2+\underline{\beta}Z_4\},\quad {\mathcal L}_{308}=\{P_0+\varepsilon T_1+\mu T_2+\underline{\alpha}Z_2+\underline{\beta}Z_4\},\\ 
{\mathcal L}_{309}&=\{P_1+\varepsilon T_1+\mu T_2+\underline{\alpha}Z_2+\underline{\beta}Z_4\},\\ 
{\mathcal L}_{310}&=\{P_0+\varepsilon P_1+\mu T_1+kT_2+\underline{\alpha}Z_2+\underline{\beta}Z_4\}{\mbox{ (where }k\neq 0)},\quad {\mathcal L}_{311}=\{Z_1+kZ_2+lZ_4\}{\mbox{ (where }k,l\neq 0)},\\ 
{\mathcal L}_{312}&=\{M_1+2M_2-M_4+\underline{\alpha}Z_1+\underline{\beta}Z_2+\underline{\gamma}Z_4\},\quad
{\mathcal L}_{313}=\{P_0+\underline{\alpha}Z_1+\underline{\beta}Z_2+\underline{\gamma}Z_4\},\\ 
{\mathcal L}_{314}&=\{P_1+\underline{\alpha}Z_1+\underline{\beta}Z_2+\underline{\gamma}Z_4\},\quad {\mathcal L}_{315}=\{P_0+\varepsilon P_1+\underline{\alpha}Z_1+\underline{\beta}Z_2+\underline{\gamma}Z_4\},\\ 
{\mathcal L}_{316}&=\{T_1+\underline{\alpha}Z_1+\underline{\beta}Z_2+\underline{\gamma}Z_4\},\quad {\mathcal L}_{317}=\{M_1+2M_2-M_4+\underline{\alpha}Z_1+\underline{\beta}Z_2+\underline{\gamma}Z_4\},\\ 
{\mathcal L}_{318}&=\{P_0+\varepsilon T_1+\underline{\alpha}Z_1+\underline{\beta}Z_2+\underline{\gamma}Z_4\},\quad
{\mathcal L}_{319}=\{P_1+\varepsilon T_1+\underline{\alpha}Z_1+\underline{\beta}Z_2+\underline{\gamma}Z_4\},\\ 
{\mathcal L}_{320}&=\{P_0+\varepsilon P_1+\mu T_1+\underline{\alpha}Z_1+\underline{\beta}Z_2+\underline{\gamma}Z_4\},\quad {\mathcal L}_{321}=\{T_2+\underline{\alpha}Z_1+\underline{\beta}Z_2+\underline{\gamma}Z_4\},\\ 
{\mathcal L}_{322}&=\{P_0+\varepsilon T_2+\underline{\alpha}Z_1+\underline{\beta}Z_2+\underline{\gamma}Z_4\},\quad {\mathcal L}_{323}=\{P_1+\varepsilon T_2+\underline{\alpha}Z_1+\underline{\beta}Z_2+\underline{\gamma}Z_4\},\\ 
{\mathcal L}_{324}&=\{P_0+\varepsilon P_1+\mu T_2+\underline{\alpha}Z_1+\underline{\beta}Z_2+\underline{\gamma}Z_4\},\quad
{\mathcal L}_{325}=\{T_1+\varepsilon T_2+\underline{\alpha}Z_1+\underline{\beta}Z_2+\underline{\gamma}Z_4\},\\ 
{\mathcal L}_{326}&=\{P_0+\varepsilon T_1+\mu T_2+\underline{\alpha}Z_1+\underline{\beta}Z_2+\underline{\gamma}Z_4\},\quad {\mathcal L}_{327}=\{P_1+\varepsilon T_1+\mu T_2+\underline{\alpha}Z_1+\underline{\beta}Z_2+\underline{\gamma}Z_4\},\\ 
{\mathcal L}_{328}&=\{P_0+\varepsilon P_1+\mu T_1+kT_2+\underline{\alpha}Z_1+\underline{\beta}Z_2+\underline{\gamma}Z_4\}{\mbox{ (where }k\neq 0)},\quad {\mathcal L}_{329}=\{Z_3+kZ_4\}{\mbox{ (where }k\neq 0)},\\ 
{\mathcal L}_{330}&=\{M_3+\underline{\alpha}Z_3+\underline{\beta}Z_4\},\quad
{\mathcal L}_{331}=\{M_1+\frac{3}{2}M_2+bM_3+\underline{\alpha}Z_3+\underline{\beta}Z_4\},\\ 
{\mathcal L}_{332}&=\{P_0+\underline{\alpha}Z_3+\underline{\beta}Z_4\},\quad {\mathcal L}_{333}=\{M_3+\varepsilon P_0+\underline{\alpha}Z_3+\underline{\beta}Z_4\},\quad {\mathcal L}_{334}=\{P_1+\underline{\alpha}Z_3+\underline{\beta}Z_4\},\\
\end{split}
\end{displaymath}
\begin{displaymath}
\begin{split}
{\mathcal L}_{335}&=\{M_3+\varepsilon P_1+\underline{\alpha}Z_3+\underline{\beta}Z_4\},\quad {\mathcal L}_{336}=\{M_3+\varepsilon P_0+\mu P_1+\underline{\alpha}Z_3+\underline{\beta}Z_4\},\\
{\mathcal L}_{337}&=\{P_0+\varepsilon P_1+\underline{\alpha}Z_3+\underline{\beta}Z_4\},\quad {\mathcal L}_{338}=\{T_1+\underline{\alpha}Z_3+\underline{\beta}Z_4\},\\ 
{\mathcal L}_{339}&=\{M_1+\frac{3}{2}M_2-\frac{1}{2}M_3+\varepsilon T_1+\underline{\alpha}Z_3+\underline{\beta}Z_4\},\quad {\mathcal L}_{340}=\{P_0+\varepsilon T_1+\underline{\alpha}Z_3+\underline{\beta}Z_4\},\\
{\mathcal L}_{341}&=\{P_1+\varepsilon T_1+\underline{\alpha}Z_3+\underline{\beta}Z_4\},\quad {\mathcal L}_{342}=\{P_0+\varepsilon P_1+\mu T_1+\underline{\alpha}Z_3+\underline{\beta}Z_4\},\\
{\mathcal L}_{343}&=\{T_2+\underline{\alpha}Z_3+\underline{\beta}Z_4\},\quad {\mathcal L}_{344}=\{M_1+\frac{3}{2}M_2+M_3+\varepsilon T_2+\underline{\alpha}Z_3+\underline{\beta}Z_4\},\\ 
{\mathcal L}_{345}&=\{P_0+\varepsilon T_2+\underline{\alpha}Z_3+\underline{\beta}Z_4\},\quad {\mathcal L}_{346}=\{P_1+\varepsilon T_2+\underline{\alpha}Z_3+\underline{\beta}Z_4\},\\ 
{\mathcal L}_{347}&=\{P_0+\varepsilon P_1+\mu T_2+\underline{\alpha}Z_3+\underline{\beta}Z_4\},\quad {\mathcal L}_{348}=\{T_1+\varepsilon T_2+\underline{\alpha}Z_3+\underline{\beta}Z_4\},\\
{\mathcal L}_{349}&=\{P_0+\varepsilon T_1+\mu T_2+\underline{\alpha}Z_3+\underline{\beta}Z_4\},\quad {\mathcal L}_{350}=\{P_1+\varepsilon T_1+\mu T_2+\underline{\alpha}Z_3+\underline{\beta}Z_4\},\\ 
{\mathcal L}_{351}&=\{P_0+\varepsilon P_1+\mu T_1+kT_2+\underline{\alpha}Z_3+\underline{\beta}Z_4\}{\mbox{ (where }k\neq 0)},\quad {\mathcal L}_{352}=\{Z_1+kZ_3+lZ_4\}{\mbox{ (where }k,l\neq 0)},\\ 
{\mathcal L}_{353}&=\{M_1+\frac{3}{2}M_2+\underline{\alpha}Z_1+\underline{\beta}Z_3+\underline{\gamma}Z_4\},\quad {\mathcal L}_{354}=\{P_0+\underline{\alpha}Z_1+\underline{\beta}Z_3+\underline{\gamma}Z_4\},\\
{\mathcal L}_{355}&=\{P_1+\underline{\alpha}Z_1+\underline{\beta}Z_3+\underline{\gamma}Z_4\},\quad {\mathcal L}_{356}=\{P_0+\varepsilon P_1+\underline{\alpha}Z_1+\underline{\beta}Z_3+\underline{\gamma}Z_4\},\\ 
{\mathcal L}_{357}&=\{T_1+\underline{\alpha}Z_1+\underline{\beta}Z_3+\underline{\gamma}Z_4\},\quad {\mathcal L}_{358}=\{P_0+\varepsilon T_1+\underline{\alpha}Z_1+\underline{\beta}Z_3+\underline{\gamma}Z_4\},\\ 
{\mathcal L}_{359}&=\{P_1+\varepsilon T_1+\underline{\alpha}Z_1+\underline{\beta}Z_3+\underline{\gamma}Z_4\},\quad {\mathcal L}_{360}=\{P_0+\varepsilon P_1+\mu T_1+\underline{\alpha}Z_1+\underline{\beta}Z_3+\underline{\gamma}Z_4\},\\
{\mathcal L}_{361}&=\{T_2+\underline{\alpha}Z_1+\underline{\beta}Z_3+\underline{\gamma}Z_4\},\quad {\mathcal L}_{362}=\{P_0+\varepsilon T_2+\underline{\alpha}Z_1+\underline{\beta}Z_3+\underline{\gamma}Z_4\},\\ 
{\mathcal L}_{363}&=\{P_1+\varepsilon T_2+\underline{\alpha}Z_1+\underline{\beta}Z_3+\underline{\gamma}Z_4\},\quad {\mathcal L}_{364}=\{P_0+\varepsilon P_1+\mu T_2+\underline{\alpha}Z_1+\underline{\beta}Z_3+\underline{\gamma}Z_4\},\\ 
{\mathcal L}_{365}&=\{T_1+\varepsilon T_2+\underline{\alpha}Z_1+\underline{\beta}Z_3+\underline{\gamma}Z_4\},\quad {\mathcal L}_{366}=\{P_0+\varepsilon T_1+\mu T_2+\underline{\alpha}Z_1+\underline{\beta}Z_3+\underline{\gamma}Z_4\},\\
{\mathcal L}_{367}&=\{P_1+\varepsilon T_1+\mu T_2+\underline{\alpha}Z_1+\underline{\beta}Z_3+\underline{\gamma}Z_4\},\\ 
{\mathcal L}_{368}&=\{P_0+\varepsilon P_1+\mu T_1+kT_2+\underline{\alpha}Z_1+\underline{\beta}Z_3+\underline{\gamma}Z_4\}{\mbox{ (where }k\neq 0)},\\ {\mathcal L}_{369}&=\{Z_2+kZ_3+lZ_4\}{\mbox{ (where }k,l\neq 0)},\quad {\mathcal L}_{370}=\{M_1+\frac{3}{2}M_2+\frac{1}{2}M_3+\underline{\alpha}Z_2+\underline{\beta}Z_3+\underline{\gamma}Z_4\},\\ 
{\mathcal L}_{371}&=\{P_0+\underline{\alpha}Z_2+\underline{\beta}Z_3+\underline{\gamma}Z_4\},\quad {\mathcal L}_{372}=\{P_1+\underline{\alpha}Z_2+\underline{\beta}Z_3+\underline{\gamma}Z_4\},\\
{\mathcal L}_{373}&=\{P_0+\varepsilon P_1+\underline{\alpha}Z_2+\underline{\beta}Z_3+\underline{\gamma}Z_4\},\quad {\mathcal L}_{374}=\{T_1+\underline{\alpha}Z_2+\underline{\beta}Z_3+\underline{\gamma}Z_4\},\\ 
{\mathcal L}_{375}&=\{P_0+\varepsilon T_1+\underline{\alpha}Z_2+\underline{\beta}Z_3+\underline{\gamma}Z_4\},\quad {\mathcal L}_{376}=\{P_1+\varepsilon T_1+\underline{\alpha}Z_2+\underline{\beta}Z_3+\underline{\gamma}Z_4\},\\ 
{\mathcal L}_{377}&=\{P_0+\varepsilon P_1+\mu T_1+\underline{\alpha}Z_2+\underline{\beta}Z_3+\underline{\gamma}Z_4\},\quad {\mathcal L}_{378}=\{T_2+\underline{\alpha}Z_2+\underline{\beta}Z_3+\underline{\gamma}Z_4\},\\
{\mathcal L}_{379}&=\{P_0+\varepsilon T_2+\underline{\alpha}Z_2+\underline{\beta}Z_3+\underline{\gamma}Z_4\},\quad {\mathcal L}_{380}=\{P_1+\varepsilon T_2+\underline{\alpha}Z_2+\underline{\beta}Z_3+\underline{\gamma}Z_4\},\\ 
{\mathcal L}_{381}&=\{P_0+\varepsilon P_1+\mu T_2+\underline{\alpha}Z_2+\underline{\beta}Z_3+\underline{\gamma}Z_4\},\quad {\mathcal L}_{382}=\{T_1+\varepsilon T_2+\underline{\alpha}Z_2+\underline{\beta}Z_3+\underline{\gamma}Z_4\},\\ 
{\mathcal L}_{383}&=\{P_0+\varepsilon T_1+\mu T_2+\underline{\alpha}Z_2+\underline{\beta}Z_3+\underline{\gamma}Z_4\},\quad {\mathcal L}_{384}=\{P_1+\varepsilon T_1+\mu T_2+\underline{\alpha}Z_2+\underline{\beta}Z_3+\underline{\gamma}Z_4\},\\
{\mathcal L}_{385}&=\{P_0+\varepsilon P_1+\mu T_1+kT_2+\underline{\alpha}Z_2+\underline{\beta}Z_3+\underline{\gamma}Z_4\}{\mbox{ (where }k\neq 0)},\\ {\mathcal L}_{386}&=\{Z_1+kZ_2+lZ_3+mZ_4\}{\mbox{ (where }k,l,m\neq 0)},\quad {\mathcal L}_{387}=\{P_0+\underline{\alpha}Z_1+\underline{\beta}Z_2+\underline{\gamma}Z_3+\underline{\delta}Z_4\},\\ 
{\mathcal L}_{388}&=\{P_1+\underline{\alpha}Z_1+\underline{\beta}Z_2+\underline{\gamma}Z_3+\underline{\delta}Z_4\},\quad {\mathcal L}_{389}=\{P_0+\varepsilon P_1+\underline{\alpha}Z_1+\underline{\beta}Z_2+\underline{\gamma}Z_3+\underline{\delta}Z_4\},\\ 
{\mathcal L}_{390}&=\{T_1+\underline{\alpha}Z_1+\underline{\beta}Z_2+\underline{\gamma}Z_3+\underline{\delta}Z_4\},\\
\end{split}
\end{displaymath}
\begin{displaymath}
\begin{split}
{\mathcal L}_{391}&=\{P_0+\varepsilon T_1+\underline{\alpha}Z_1+\underline{\beta}Z_2+\underline{\gamma}Z_3+\underline{\delta}Z_4\},\quad {\mathcal L}_{392}=\{P_1+\varepsilon T_1+\underline{\alpha}Z_1+\underline{\beta}Z_2+\underline{\gamma}Z_3+\underline{\delta}Z_4\},\\ 
{\mathcal L}_{393}&=\{P_0+\varepsilon P_1+\mu T_1+\underline{\alpha}Z_1+\underline{\beta}Z_2+\underline{\gamma}Z_3+\underline{\delta}Z_4\},\quad {\mathcal L}_{394}=\{T_2+\underline{\alpha}Z_1+\underline{\beta}Z_2+\underline{\gamma}Z_3+\underline{\delta}Z_4\},\\ 
{\mathcal L}_{395}&=\{P_0+\varepsilon T_2+\underline{\alpha}Z_1+\underline{\beta}Z_2+\underline{\gamma}Z_3+\underline{\delta}Z_4\},\quad {\mathcal L}_{396}=\{P_1+\varepsilon T_2+\underline{\alpha}Z_1+\underline{\beta}Z_2+\underline{\gamma}Z_3+\underline{\delta}Z_4\},\\
{\mathcal L}_{397}&=\{P_0+\varepsilon P_1+\mu T_2+\underline{\alpha}Z_1+\underline{\beta}Z_2+\underline{\gamma}Z_3+\underline{\delta}Z_4\},\\ 
{\mathcal L}_{398}&=\{T_1+\varepsilon T_2+\underline{\alpha}Z_1+\underline{\beta}Z_2+\underline{\gamma}Z_3+\underline{\delta}Z_4\},\\ 
{\mathcal L}_{399}&=\{P_0+\varepsilon T_1+\mu T_2+\underline{\alpha}Z_1+\underline{\beta}Z_2+\underline{\gamma}Z_3+\underline{\delta}Z_4\},\\ 
{\mathcal L}_{400}&=\{P_1+\varepsilon T_1+\mu T_2+\underline{\alpha}Z_1+\underline{\beta}Z_2+\underline{\gamma}Z_3+\underline{\delta}Z_4\},\\ 
{\mathcal L}_{401}&=\{P_0+\varepsilon P_1+\mu T_1+kT_2+\underline{\alpha}Z_1+\underline{\beta}Z_2+\underline{\gamma}Z_3+\underline{\delta}Z_4\}{\mbox{ (where }k\neq 0)}.
\end{split}
\end{displaymath}

{}

\label{lastpage}
\end{document}